\documentclass[prc,showpacs,preprintnumbers,amsmath,amssymb,floatfix]{revtex4}
\usepackage{graphicx}
\usepackage{dcolumn}
\usepackage{amsmath}
\renewcommand{\vec}[1]{\boldsymbol{#1}}

\begin{document} 
\title{\Large 
Shape mixing in the $A\approx 70$ region} 
\author{Daniel Almehed}
\email{D.Almehed@surrey.ac.uk}
\affiliation{Department of Physics, University of Surrey, Guildford GU2 7XH, 
United Kingdom} 
\author{Niels R.~Walet} 
\email{Niels.Walet@manchester.ac.uk}
\affiliation{School of Physics and
Astronomy, The University of Manchester, P.O. Box 88, Manchester M60
1QD, United~Kingdom} 
 
\begin{abstract}
We investigate the mixing of different shapes in the $A\approx70$ region
using the adiabatic self-consistent collective method in a rotating nuclei.
The calculation is done for $^{68}$Se and $^{72-78}$Kr which are known to 
show  oblate-prolate shape coexistence at low angular momentum. A
pairing-plus-quadrupole Hamiltonian, which is known to reproduce reproduce 
the structure nuclei in this mass region, is used. We calculate the collective 
path between 
the oblate and prolate minima and discuss how the  collective behaviour of 
the system changes with increasing angular momentum. 
\end{abstract} 
\pacs{21.60.-n, 21.60.Jz}
\maketitle 

\section{Introduction}

The question ``what is the shape'' of a nucleus is an old one, and
probably dates back to the introduction of Bohr's liquid drop model
\cite{BK37}.  Modern nuclear physics has tried to find answers to this question
in many areas of the table of isotopes. One area of extreme
experimental interest is the $N\approx Z$ region for intermediate
mass, where all the simple models tell us nuclear dynamics rapidly
changes with proton and neutron number. Such proton-rich nuclei are
also important in the RP-process of nucleo-synthesis, and are of
interest since they are only accessible with modern experimental
facilities.

Mean-field methods are one way to give microscopic meaning to the
concept of nuclear shape. They have been used quite extensively in
nuclear physics \cite{RS80}, since they give a good starting point for
the description of many but the lightest of nuclei. There are a large
number of improvements one can make, either at a single minimum in the
mean-field energy (random phase phase approximation, projection
techniques, quasi-particle expansions), or in the case that a single
minimum is insufficient (time-dependent Hartree-Fock and related
methods, Hill-Wheeler method, etc.).

In this work we shall apply the theory of large-amplitude
collective motion \cite{DK00}, which is a method that 
tries to describe, in a self-consistent fashion, 
a few-dimensional surface in which most of the low-energy
dynamics takes place. There are
many related methods in use in many areas of physics, see e.g.
\cite{To98}.
Our method is just one of a large set of techniques
applied to this problem in nuclear physics \cite{DK00}. 

Even though it is derived in a very different way, it shows great
similarity to the Hill-Wheeler method, which relies on the constrained
Hartree-Fock Bogoliubov (CHFB) technique. Here a set of Slater
determinants determined by a small set of judiciously chosen one-body
constraints is used to map out what is presumably a dominant part of
the space of all configurations. The analogy with the technique of
generating states for specified angular momentum by imposing
constraints (cranking) leads to the name ``generalised cranking''.
The one-body constraints usually consist of a few carefully chosen
multipole, or particle-hole, operators, as well as a few generalised
pairing, or particle-particle, ones.  For large scale realistic problems
such as the description of nuclear fission the number of generalised
cranking operators needed in order to make a realistic calculation can
become very large.  There is also no reason to limit the constraints
to the standard choices; other degrees of freedom, especially those
involving spin-orbit interactions might also be important. A more
satisfactory method should allow the cranking operators to be
determined by the nuclear collective dynamics itself.

One such approach, followed in this paper and set out in detail in the
review paper Ref.~\cite{DK00} as well as in Ref.~\cite{AW04} (a
similar approach, plus relevant references, can be found in
Refs.~\cite{MN00,KN03,KN04,KN05}), leads to a very well-defined
approach, which can in principle be solved knowing the Hamiltonian and
model space.  To find the adiabatic collective path we use the local
harmonic approximation (LHA).  It consists of a constrained mean-field
problem that needs to be solved together with a local random phase
approximation (RPA), which determines the constraining operator. This
approach lacks practicality, since the size of the RPA problem is, for
a system with pairing, proportional to the size of the single-particle
space squared. Even though enormous matrices can routinely be
diagonalised on modern computer systems, the algorithm requires
repeated diagonalisation of such a matrix, which makes an
implementation in realistic calculations prohibitively time consuming.
A practical solution to this problem was given in
Ref.~\cite{NW99,AW03,AW04}.  It consist of solving the RPA in a
limited basis of state dependent operators.  This was shown to give
excellent results for low lying RPA solutions in the minimum as well
as along the collective path. Our Japanese colleagues have developed
an interesting alternative based on the full RPA for separable forces,
which can be solved by dispersion relations \cite{KN03,KN04,KN05}.

In this paper we set out to study the the collective motion in rotating 
nuclei. The method is a generalised version of the method presented in 
Ref.~\cite{AW04} where we study collective path at a given angular 
momentum~\cite{AW04b}.
Experimental results show a large variety of shape-coexistence
phenomena at low angular momentum in the mass-region around
$^{72}$Kr~\cite{FB00,BM03,CR97,BK99,KC05}. This is therefore a good testing 
ground for large amplitude collective motion calculations in rotating nuclei.

\section{Formalism}
\label{sec:Formalism}
Details of the formalism can be found in our previous work,
specifically Refs.~\cite{DK00,AW04}.  It is based on time-dependent
mean field theory, and the fact that this is a form of classical
Hamiltonian dynamics.  The issue of selecting collective coordinates
can then be separated in two steps, the determination of the
Hamiltonian governing the classical mechanics and the approximate
decoupling of a few modes in classical mechanics.  To facilitate the
second task, we usually make the additional assumption of slow motion,
called the adiabatic approximation.  In that limit, it sufficed to
expand the Hamiltonian to second order in momenta, and we have a
problem that can be solved in closed form.  The solution, as
succinctly summarised below, can be stated without any 
reference to the original nuclear many-body problem and the choice of
the interaction. The alternative method set out in Ref.~\cite{MN00} looks
rather different, and relies heavily on the use of Slater determinants.
As discussed in the appendix, it is very close to our work and only  differs
in small details
.
\subsection{Local harmonic approximation for the collective path in the 
	adiabatic limit\label{sec:LHA}}
By parameterising the expectation value of the quantum-Hamiltonian
with a set of classical coordinates, which corresponds to a
classical boson mapping of the density matrix \cite{DK00},
we can parametrise the Hamiltonian locally in terms of a set
of coordinates and momenta. Clearly, we could also
work in the language of differential forms, and avoid specifying
coordinates, and coordinate frames that are chosen differently
at different points \cite{AM88}. Such elegance would hide
the technical aspects of obtaining a suitable algorithm.

We assume a classical Hamiltonian depending on a set of real canonical
coordinates, $\xi^\alpha$ ($\alpha=1,\ldots ,N$), and conjugate
momenta, $\pi_\beta$ ($\beta=1,\ldots ,N$).
The potential $V(\xi)={\cal H}(\pi=0)$ and the mass matrix $B^{\alpha \beta}$
are given by an expansion of ${\cal H}(\xi,\pi)$ in powers of $\pi$ in zeroth
and second order, respectively,
\begin{equation}
	\label{eq:H2}
	{\cal H}(\xi,\pi) = V(\xi) + \frac{1}{2} B^{\alpha \beta} \pi_\alpha 
	\pi_\beta + {\cal O}(\pi^4) .
\end{equation}
Terms of higher order (such as $\pi^4$) are supposed to be negligible.
The central part in our approach to large amplitude motion is a search
for collective (and non-collective) coordinates $q^\mu$ which are
obtained by an invertible point transformation of the original
coordinates $\xi^\alpha$, preserving the quadratic truncation of the
momentum dependence of the Hamiltonian,  by
\begin{equation}
	\label{eq:fg1}
	q^\mu = f^\mu(\xi), \quad \xi^\alpha = g^\alpha(q) \quad \left( 
	\mu , \alpha =  1,\ldots ,n \right) ,
\end{equation}
and the corresponding transformation relations for the momenta
$p_\mu$ and $\pi_\alpha$,
\begin{equation}
	\label{eq:fg2}
	p_\mu = g^\alpha_{,\mu}\pi_\alpha, \quad \pi_\alpha = 
	f^\mu_{,\alpha} p_\mu \quad,
\end{equation}
where we use a standard notation for the derivatives,
$g^\alpha_{,\mu} \equiv \frac{\partial}{\partial q^\mu} g^\alpha$ and 
$f^\mu_{,\alpha} \equiv \frac{\partial}{\partial \xi^\alpha} f^\mu$.
The adiabatic Hamiltonian, Eq.~(\ref{eq:H2}), is then 
transformed into
\begin{equation}
	\label{eq:H3}
	\bar{{\cal H}}(q,p) = \bar{V}(q) + \frac{1}{2} \bar{B}^{\mu \nu} 
	p_\mu p_\nu + {\cal O}(p^4) 
\end{equation}
in the new coordinates. The new coordinates $q^\mu$ are  to be divided
into three  categories: the collective coordinate $q^1$, the
zero-mode coordinates $q^I$, $I=2,\ldots,M+1$, which describe motions that
do not change the energy and finally the non-collective
coordinates $q^a$, $a=M+2, \ldots ,n$. 

The collective coordinate is determined by means of the
solution to the local harmonic approach, 
which consists of a set of self-consistent
equations.  These are:
\begin{enumerate}
\item The force equations
\begin{equation}
	\label{eq:force1}
	\bar{{\cal H}}_{,\alpha} = \Lambda f_{,\alpha} + \Lambda_I f^I_{,\alpha} ,
\end{equation}
where $f^I$ are the zero-modes (also called Nambu-Goldstone or
spurious modes) and $\Lambda_I$ represents a set of Lagrange
multipliers (which in nuclear physics are sometimes called cranking
parameters).  $\Lambda$ is a Lagrange multiplier for the collective
mode, stabilising the system away from equilibrium (we shall often
call it  the generalised cranking parameter).
\item The local RPA equation
\begin{equation}
	\label{eq:localRPA1}
	\bar{V}_{;\alpha \gamma} B^{\gamma \beta} f_{,\beta} = 
	\left( \hbar \Omega \right)^2 f_{,\alpha} ,
\end{equation}
where the covariant derivative $V_{;\alpha \beta}$ is defined in the
usual way ($V_{,\alpha \beta}=(V_{,\alpha})_{, \beta}$).
Zero modes correspond to zero eigenvalues of the RPA. In principle
great care needs to be taken to have zero modes behave correctly away
from equilibrium. The symplectic RPA~\cite{DK00} is the correct way to
do so; unfortunately it is rather cumbersome. Below we will discuss an 
alternative approach to remove the admixture of spurious solution in an 
approximate scheme. 
\end{enumerate}
The collective path is found by solving Eqs.~(\ref{eq:force1})
and~(\ref{eq:localRPA1}) self-consistently, i.e., we look for a path
consisting of a series of points where the lowest non-spurious
eigenvector of the local RPA equations also fulfils the force
condition. The motion along this line does not necessarily decouple 
from other degrees of freedom, but we have a ``decoupling measure'' $D$,
that measures how good or bad this coupling is. As discussed
in Ref.~\cite{DK00}, it measures the change of non-collective coordinates
for a change in the collective one, which must be zero for exact decoupling.
In the minimum of the potential the spurious solutions decouple from
the other collective and non-collective solutions, and spurious admixtures
are absent. This is not true for other points along the collective path, 
especially when approximations are made~\cite{Yo04}, see the further discussion
in Sec.~\ref{sec:spur}.

\subsection{Pairing+quadrupole model}
We apply the LHA to the pairing+quadrupole Hamiltonian as described 
in~\cite{BK68}. With a constraint on both angular momentum and particle 
numbers the Hamiltonian can be written as
\begin{eqnarray}
	\label{eq:HPQ1}
	H' &=& H - \sum_{\tau=n,p} \mu_\tau N_\tau - \omega J , \\
	\label{eq:HPQ2}
	H &=& \sum_l \epsilon_l c^\dagger_k c_k - \sum_{\tau=n,p} 
	\frac{G_\tau}{2} \left( P^\dagger_\tau P_\tau + P_\tau P^\dagger_\tau 
	\right) - \frac{\kappa}{2} \sum_{M=-2}^2 Q^\dagger_{2M} Q_{2M} , 
\end{eqnarray}
where $\epsilon_l$ are spherical single-particle energies, $N_\tau$ are the 
particle number operators, $J$ is the angular momentum operator, $Q_{2M}$ is 
the dimensionless quadrupole operator
\begin{equation}
	\label{eq:Q1}
	Q_{2M} = \frac{1}{\sqrt{2} b^2_0} \sum_{kl} \left< k \right| r^2 Y_{2M}
	\left| l \right> c^\dagger_k c_l ,
\end{equation}
where $b_0=1/\sqrt{\omega_0}$ is the standard oscillator length
and $P^\dagger_\tau$ is the (dimensionless) pairing operator
\begin{equation}
	\label{eq:P1}
	P^\dagger_\tau = \sum_{k>0} c^\dagger_k c^\dagger_{-k}\quad.
\end{equation}
In Eqs.~(\ref{eq:HPQ1}--\ref{eq:P1}) $c^\dagger_k$ ($c_k$) are the
single particle creation (destruction) operators, where $k$ labels a
set of quantum numbers. The Hamiltonian (\ref{eq:HPQ1}) is treated in
the Hartree-Bogoliubov approximation and it has been shown that at the
minimum the local RPA for this Hamiltonian is equivalent to the
quasi-particle RPA.

The spherical single-particle energies are calculated from a modified
spherical oscillator~\cite{NR95}. The interaction strengths (see 
Table~\ref{tab:KappaG}) are chosen to give realistic values of the 
deformation and pairing gaps in the ground states. 
\begin{table}
\caption{The interaction strengths used in the calculations.}
\label{tab:KappaG} 
\begin{tabular}{|r|ccc|}
\hline
  & $\kappa$ [MeV] & $G_n$ [MeV] & $G_p$ [MeV] \\ 
\hline
 $^{68}$Se & 0.09894 & 0.3437 & 0.3441 \\ 
 $^{72}$Kr & 0.091 & 0.3469 & 0.3514 \\  
 $^{74}$Kr & 0.08643 & 0.3333 & 0.3357 \\  
 $^{76}$Kr & 0.08473 & 0.3204 & 0.3326 \\  
 $^{78}$Kr & 0.08452 & 0.3077 & 0.3287 \\ 
\hline
\end{tabular}
\end{table} 
Our model space consists of two major shells. We
follow~\cite{BK68} and multiply each quadrupole matrix element with a
$N$-dependent scaling factor. To achieve the same root-mean-square
radii for protons and neutrons different harmonic oscillator
frequencies are adopted for each type of nucleons~\cite{BK68}.  The
deformation parameters $\epsilon_2$ and $\gamma$ are calculated from
the expectation values of the $Q_0$ and $Q_2$ operators as
\begin{eqnarray}
  \label{eq:eps2}
  \epsilon_2 &=& \frac{3}{2} \frac{\kappa}{\hbar \omega_0} 
  \sqrt{\left<Q_0\right>^2 + \left<Q_2\right>^2 },\\
  \label{eq:gamma}
  \gamma &=& \tan^{-1}\left(\frac{\left<Q_2\right>}{\left<Q_0\right>}\right).
\end{eqnarray} 

\subsection{Removing spurious admixtures in RPA\label{sec:spur}}

Two of the main difficulties of applying the LHA method to realistic
nuclear problems are the effort required in diagonalising the
large-dimensional RPA matrix repeatedly within the double iterative
process and the problem of spurious admixtures into the collective path. 
To limit the computational effort we use the method
presented in Ref.~\cite{NW99,AW04} to reduce the size of the RPA matrix.
There it was shown that the RPA equation can be solved with good
accuracy by assuming that the RPA eigenvectors can be described as a
linear combination of a small number of state-dependent one-body
operators. The same set of operators turned our to give good bases for
solving the RPA problem also at finite angular momentum~\cite{AW04b}.
To avoid problem with spurious admixture in our collective coordinate me 
remove the spurious admixture from our state dependent basis set. 

We select a small number of state dependent one-body operators 
$F^{(k)}$, $k=1,\ldots,n$, assuming that the RPA eigenvectors can be 
approximated as linear combinations of the $F^{(k)}$. The approximate RPA 
vector $\bar{f}_{,\alpha}$ is then given by 
\begin{equation}
	f_{,\alpha} \approx \bar{f}_{,\alpha} = \sum_{k=1}^{n} b_k 
		{\cal F}^{(k)}_{,\alpha}
	\label{eq:f1}
\end{equation}
where ${\cal F}^{(k)}$ is the expectation value of $F^{(k)}$.
Any spurious admixture to this basis RPA vector is removed by enforcing that the
vector ${\cal F}^{(k)}$ is perpendicular to the spurious particle number and 
angular momentum operators. This new basis set $\tilde{\cal F}^{(k)}$ is given by
\begin{equation}
  \tilde{{\cal F}}^{(k)} = {\cal F}^{(k)} - \sum_i d_i^k O^{(i)}\quad,
  \label{eq:spor1}
\end{equation}
where $O^{(i)}$ are the particle number and angular momentum operators. 
The coefficients $d_i^k$ can be found by enforcing the condition
\begin{equation}
  \tilde{{\cal F}}^{(k)} \cdot O^{(j)} = 0\quad,
  \label{eq:spor2}
\end{equation}
which leads to a system of linear equations which can be solved for
$d_i^k$. Even though the removal of the spurious admixture is of
principal importance it turns out to only give a small change to the
collective path for the example of $^{72}$Kr as studied in
Ref.~\cite{AW04b}.

\subsection{Schr\"{o}dinger equation on the collective path}
\label{sec:SE}

After having made a semi-classical approximation, which leads to a
classical Hamiltonian, we need to remember that we are studying a
quantum system. The standard technique to deal with this is to treat the
classical Hamiltonian as a quantum one, and to calculate the
eigenfunctions and energies. 
One must include the zero modes when quantising the Hamiltonian, since
they describe rotational and other excitations. Quantisation of the
Hamiltonian in a metric coordinate space turns the kinetic energy into a
Laplace-Beltrami operator (see, e.g., Ref.~\cite{AM88}) in the
relevant space and 
the collective Schr\"{o}dinger equation can then be written as
\begin{equation}
	\label{eq:SE1}
	H(\vec{ X}) \Psi(\vec{ X}) = E \Psi(\vec{ X})
\end{equation}
where $\vec{X}$ represent both the collective coordinate Q and the
spurious coordinates. 

Since the potential and the masses are independent of the zero-mode coordinates
the wave-function $\Psi$ can be separated into various pieces,
\begin{equation}
  \label{eq:wf1}
  \Psi(Q,\phi_N,\phi_P,\Omega) =  g^{-1/4} U(Q) \frac{1}{\sqrt{2 \pi}}e^{i m \phi_N} 
  \frac{1}{\sqrt{2 \pi}}e^{i k \phi_P} D^I_{MK}(\Omega)^*
\end{equation}
where $m$ and $k$ are the quantum numbers for neutron and proton
pairing rotation, and $I,M,K$ are the usual rotator quantum
numbers. We shall be looking at ground states (band-heads) only, and
therefore we shall now use $I=M=K=0$, and since pairing rotational excitation 
corresponds to a change in particle number, we shall be use $m=k=0$ as well. 
Using (\ref{eq:wf1}) to separate
variables, the Schr\"{o}dinger equation~(\ref{eq:SE1}) can be written as
\begin{equation}
	\label{eq:SE2}
	-g^{-1/4} \frac{\partial}{\partial Q} 
	\left( g^{1/2} \frac{\partial }{\partial Q} g^{-1/4}U(Q)\right) 
	+V(Q) U(Q)
	= E U(Q) .
\end{equation}

Since we wish the wave function $\Psi$ to be normalisable we require
it to be finite, and we must then insist that $U(Q)$ goes to zero
when $g$ does. In the present work that only occurs when either of the
pairing gaps collapses and thus $B_{\phi_{P,N}}=0$ is zero, and we
shall ignore the rotational moments of inertia, which do not change
very quickly.  Below we solve Eq.~(\ref{eq:SE2}) on a grid with the
boundary condition that $U(Q_{\rm max})=U(Q_{\rm min})=0$. At points
where $B_\phi=0$ the condition $U=0$ holds exactly; for other cases
applying this boundary condition will only give an upper limit on
energy. 

\section{Results}
\label{sec:Results}

We have performed calculations for a chain of Krypton isotopes in the
 $A=70$ region, to study the effect of changing neutron number on
the behaviour of these nuclei. To compare with previous work \cite{KN04,KN05},
we have also studied the nucleus $^{68}$Se, which will be seen below to be 
behave in a rather similar way as some of the Krypton nuclei.

\subsection{$^{68}$Se}
The nucleus $^{68}$Se is of key interest since it has equal numbers of
protons and neutrons. It has also been investigated using similar
methods for $J=0^+$ by Matsuo {\em et al.}~\cite{KN03,KN04}.
Preliminary results of our investigation has been published in~\cite{AW05}.

\begin{figure}
  \centerline{\includegraphics[clip,width=10cm]{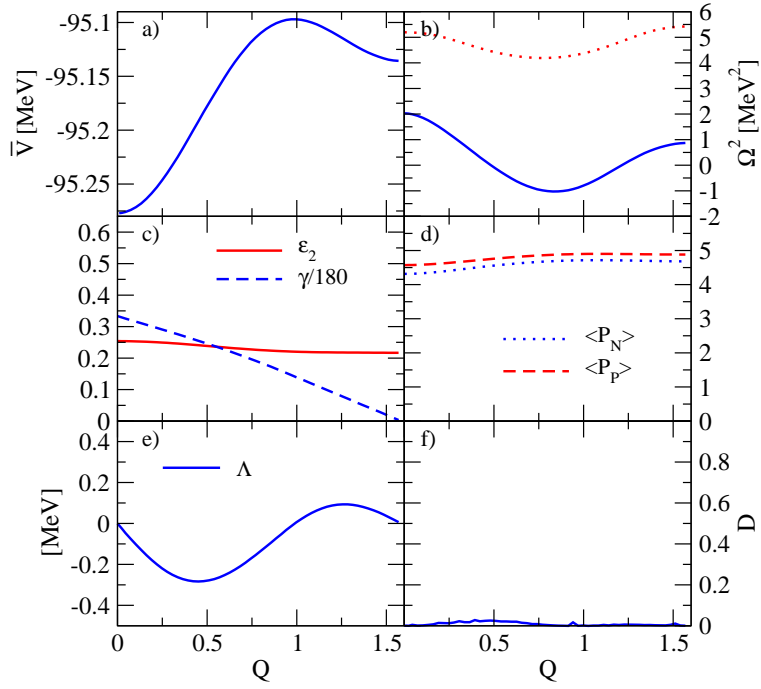}}
  \caption{Large amplitude collective motion in $^{68}$Se at $I=0$ 
  following the lowest RPA solution at equilibrium. 
  a) Energy along the collective path.  b)
  The square of the lowest RPA frequencies.  c) Deformation $\epsilon_2$ and 
  $\gamma /180^\circ$. d) The dimensionless pairing
  operators $\left< P_\tau \right>$.  e) The cranking parameter
  $\Lambda$.  f) The decoupling measure, $D$. }
  \label{fig:se68I0}
\end{figure}

We start by examining the ground state and excited $0^+$ states in
$^{68}$Se where experiment suggests prolate-oblate shape coexistence
~\cite{FB00}. We find a collective path going from the oblate minimum
over a triaxial energy maximum into a prolate secondary minimum, see
Fig.~\ref{fig:se68I0}.
\begin{figure}
  \centerline{\includegraphics[clip,width=5cm]{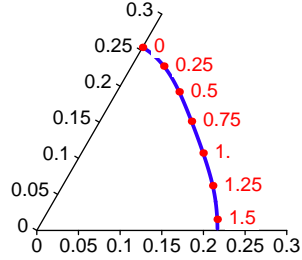}}
  \caption{ 
  The deformation parameters in the $\epsilon_2-\gamma$ plane for $^{68}$Se 
  at $I=0$. Labels denote the value of the collective coordinate.}
  \label{fig:se68I0bg}
\end{figure}
We see that the main change in the structure along the collective path
is a almost linear change in the $\gamma$-deformation. The energy
difference and the potential barrier between the two minima is very
small, less then $200$ keV. We can therefore expect to see a large
mixing of different shapes.  The collective path can be mirrored
around prolate and oblate minimum due to the time reversal
symmetry. If we were to continue the path towards negative values of
the collective coordinate $Q$ and towards larger $Q$ the potential
would repeat it self in a cyclic manner.  This collective path is very
similar to what was recently found in Ref.~\cite{KN04}. The smallness
of the decoupling measure $D$ gives us reason to believe  that
we can obtain a very good description of the ground state in this way.

\begin{figure}
  \centerline{\includegraphics[clip,width=8.0cm]{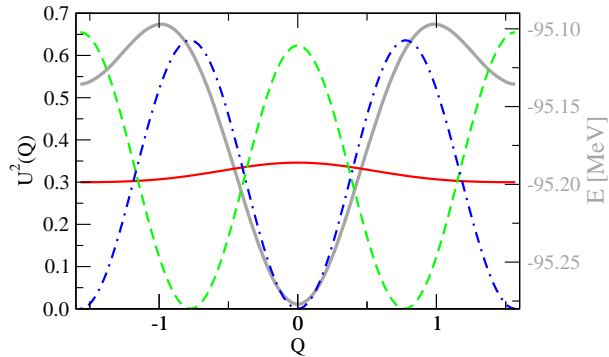}} 
  \caption{The  wave function for the large amplitude collective motion 
	in $^{68}$Se at $I=0$, following the lowest RPA solution.  The 
	first three $0^+$ eigen  functions are shown (as red solid, green
dashed and blue dashed-dotted line, respectively).
	These wave functions are periodic around $Q=0$ and $\pm 1.57$.
	The grey curve (with the scale on the 
	right-hand side) shows the collective potential energy for reference. }
  \label{fig:se68I01D}
\end{figure}

We have solved the collective Schr\"{o}dinger equation, see Sec.~\ref{sec:SE}. 
The cyclic nature of the collective path
gives   condition for the radial part of the
collective wave function, $U$, that $U(Q) = U(-Q)$. In
Fig.~\ref{fig:se68I01D} we can see that the ground state wave function
is spread out almost uniformly along the collective path.
The ground state is thus neither prolate nor oblate, which should not
surprise us since this nucleus is very $\gamma$-soft.  This results in
two (almost) degenerate excited states at 2 MeV excitation.  This
result is similar to those obtained by solving the Schr\"{o}dinger
equation on a circle with constant potential. There 
the ground state is a constant wave function, and the 
first two excited wave functions, $\cos\phi$ and $\sin\phi$,
are degenerate, and the energy spacing only depends
on the circumference of the circle.

\begin{figure}
\centerline{\includegraphics[clip,width=10cm]{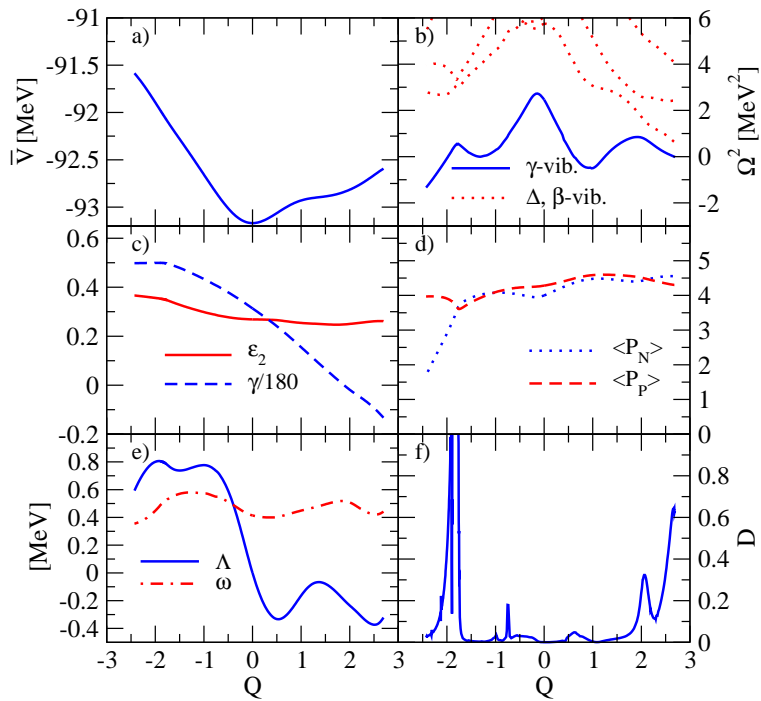}}
  \caption{Large amplitude collective motion in $^{68}$Se at $I=2$ 
  following the lowest RPA
  solution at equilibrium. a) Energy along the collective path.  b)
  The square of the lowest RPA frequencies.  c) Deformation $\epsilon_2$ and 
  $\gamma /180^\circ$. d) The dimensionless pairing
  operators $\left< P_\tau \right>$.  e) The cranking parameter
  $\Lambda$ and the rotational frequency $\omega$.  f) The decoupling
  measure, $D$. }
  \label{fig:se68I2}
\end{figure}

At finite angular momentum the behaviour gets somewhat more
complicated, as can be seen in Fig.~\ref{fig:se68I2} for $I=2$. The
potential is no longer symmetric around the minimum and for positive
values of the collective coordinate the path goes through the triaxial
plane into a very shallow prolate minimum. The collective path then
continues towards negative $\gamma$-deformation. The path ends due to
an avoided crossing with a higher lying RPA mode, at which point we
are no longer able to find a stable solution. This avoided crossing
indicates a need to include more than one collective coordinate.

This is a generic feature of the method; whenever two modes come
closely together, and thus mix strongly, we need collective
coordinates for both modes to describe collective motion. Fortunately,
the decoupling measure gives us a clear indication at what point of
the collective surface this is the case. States with wave functions
that are not sensitive to the are of large coupling should be
unaffected.

\begin{figure}
\centerline{\includegraphics[clip,width=6cm]{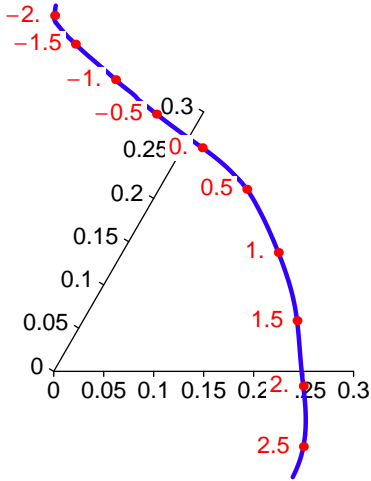}}
  \caption{The deformation parameters in the $\epsilon_2-\gamma$ plane for 
    $^{68}$Se at $I=2$. Labels denote the value of the collective coordinate.}
  \label{fig:se68I2bg}
\end{figure}
For negative $Q$ we have a linear increase in $\gamma$-deformation
until $\gamma$ reaches $90^\circ$ and $Q\approx -1.7$, where there is a
sudden change in the nature of the collective path as shown by drastic
reduction in the neutron pair-field. This is accompanied with a slow
increase of the $\epsilon_2$-deformation with $\gamma$ fixed at $90^\circ$.
Once again the size of the decoupling measure tells us that here
our single-coordinate approximation breaks down. We can
thus not reliable calculate any state that has a sizable weight near $Q=-2$.

\begin{figure}
  \centerline{\includegraphics[clip,width=8.0cm]{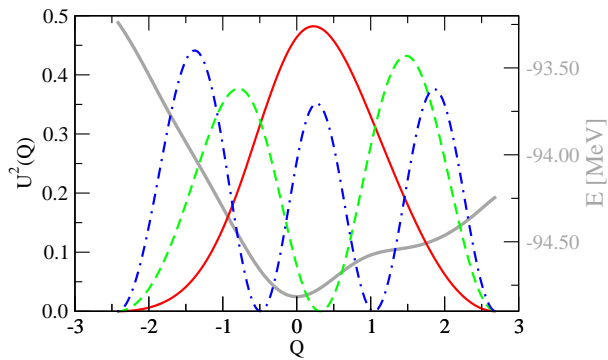}}
  \caption{The  wave function for the large amplitude collective
  motion in $^{68}$Se at $I=2$, following the lowest RPA solution. The
  first three $2^+$ eigen functions are shown (as red solid, green
  dashed and blue dashed-dotted line, respectively). The grey curve
  (with the scale on the right-hand side) shows the collective
  potential energy for reference.} \label{fig:se68I21D}
\end{figure}

After solving the collective Schr\"odinger equation, we find in
Fig.~\ref{fig:se68I21D} that the wave function of the lowest $2^+$
state is concentrated around the oblate minimum with the tail reaching
into triaxial and prolate shapes, but with little or no weight where $D$
is large.
The first excited state has two peaks in the collective wave function
corresponding to a superposition of prolate and triaxial deformations,
and is less reliable due to some small weight where $D$ is large.
It is only the third excited  $2^+$ state that appears to suffer strongly
from weighting where $D$ is appreciable.

\begin{table}
\caption{The deformation for the collective states as well as for the mean field 
	at $I=0$ and $2$ for $^{68}$Se.}
\label{tab:se68def} 
\begin{tabular}{|r|cccccccc|}
\hline
 $I_n$ & $0_1$ & $0_2$ & $0_3$ & $2_1$ & $2_2$ & $2_3$ & $0_{\rm MFG}$ & $2_{\rm MF}$\\ 
\hline
 $\epsilon_2$ & 0.211& 0.216& 0.224& 0.250& 0.229& 0.234 & 0.254 & 0.269\\ 
 $\gamma$ & 25.6 & 27.5 & 26.1 & 48.1 & 46.5 & 54.5 & 60.0 & 56.4 \\ 
\hline
\end{tabular}
\end{table} 
The average deformation for the ground state and low-lying excited
states are shown in Table~\ref{tab:se68def} together with the
mean-field deformation at $I=0$ and $2$. We conclude that
mean-field and average values can be substantially different
in these nuclei; also the difference between $0^+$ and $2^+$ states
is interesting, with the $2^+$ states much closer to purely oblate than
the $0^+$ states.

\begin{figure}
  \centerline{\includegraphics[clip,width=8cm]{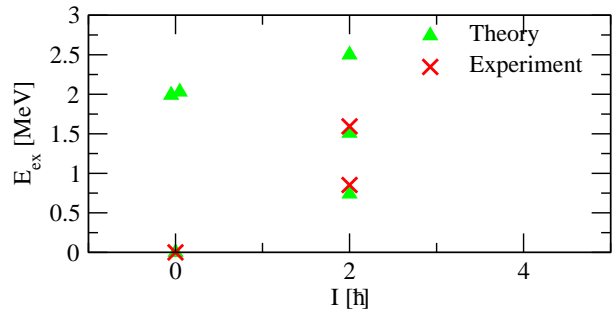}}
  \caption{Calculated and experimental~\cite{FB00} excitation spectrum for 
    $^{68}$Se as a function of angular momentum.}
  \label{fig:se68EI}
\end{figure}

If we summarise these results, and compare to the experimental
spectra, Fig.~\ref{fig:se68EI} we get some interesting results.
 The decoupling measure is small, and up to the lack of
fine-tuning of our parameters to individual nuclei, we do expect very
good results. Thus we predict a pair of almost degenerate $0^+$ states
at about 2 MeV. The results for the $2+$ states are probably better
than we could have expected: the first of these states has little
sensitivity to the large decoupling measure at $Q=-2$, but the second
has some. The third state must be thought of as totally unreliable.

\subsection{$^{72}$Kr}
\begin{figure}
\centerline{\includegraphics[clip,width=10cm]{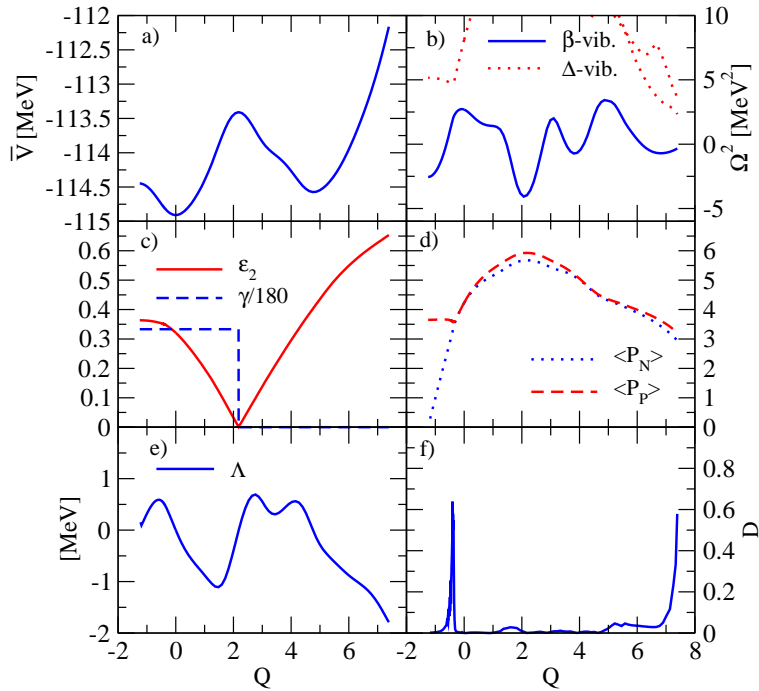}}
  \caption{Large amplitude collective motion in $^{72}$Kr with axial
  symmetry (following the $\beta$-vibration, the lowest RPA solution
  at equilibrium). a) Energy along the collective path.  b) The square
  of the lowest RPA frequencies. c) Deformation $\epsilon_2$ and
  $\gamma /180^\circ$.  d) The dimensionless pairing operators $\left<
  P_\tau \right>$.  e) The cranking parameter $\Lambda$.  f) The
  decoupling measure, $D$.}  \label{fig:kr72I0}
\end{figure}

This nucleus has been studied in detail in Ref.~\cite{AW04b}.
We again start by examining the $0^+$ states where a
prolate-oblate shape coexistence has been established
experimentally \cite{BM03}. We find a collective path going from the oblate
minimum through a spherical energy saddle point into a prolate
secondary minimum, continuing towards larger deformation, see
Fig.~\ref{fig:kr72I0}. At $Q\approx 7.5$ there is an avoided crossing
between the lowest RPA mode, which we are following, and two higher
lying modes that are of pairing-vibrational character. This can also
bee seen in Fig.~\ref{fig:kr72I0} as a rapid increase in the
decoupling measure. 
For negative $Q$, at large oblate deformation we see a collapse in
neutron pairing, which gives a natural end to the path.
Nonetheless, even at this end of the path there is an obvious avoided crossing,
which suggests that the one dimensional approximation is only of limited value
for this nucleus. Fortunately, the maximal value for $D$ of about 0.6
suggests, based on past experience, that the influence of any second coordinate
is small.

\begin{figure}
  \centerline{\includegraphics[clip,width=8.0cm]{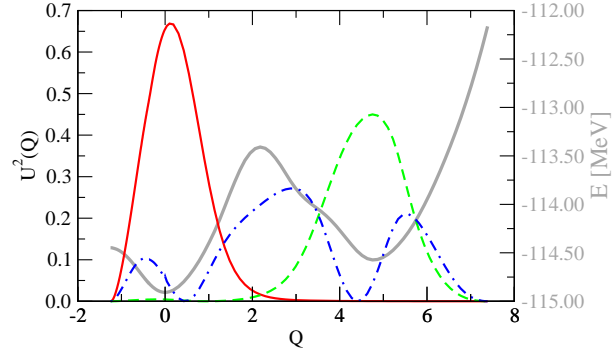}} 
  \caption{The  wave function for the large amplitude collective motion 
	in $^{72}$Kr, following the lowest RPA solution. The 
	first three $0^+$ eigen  functions are shown (as red solid, green
	dashed and blue dashed-dotted line, respectively).
	These wave functions are periodic around $Q=0$ and $\pm 1.57$.
	The grey curve (with the scale on the 
	right-hand side) shows the collective potential energy for reference. }
  \label{fig:kr72I03D}
\end{figure}

In Fig.~\ref{fig:kr72I03D} we plot the collective wave functions and can
see that the ground state wave function is concentrated in the oblate
minimum. The collective wave function has a substantial spread along
the collective path and is skewed towards the spherical state due to
the collapse in the neutron pair-field. Unfortunately, this has
substantial weight in the area of the avoided crossing.
The first excited state has its major component in the prolate minimum
with a small component in the oblate minimum. The prolate peak in the
wave function for the first excited state is broader and more
symmetric then for the ground state. The third $I=0$ state is
approximate spherical but has substantial oblate as well as prolate
components.

At finite rotational frequency the collective path will no longer go
through the spherical state, due to the effect of the rotational term
(it is energetically unfavourable to generate angular momentum in a
spherical state). Instead the oblate and prolate minim are connected
through a path that goes through the triaxial plane. Due to problems
with level crossings we have to start in both the prolate and oblate
minimum, and connect these together to find the complete collective
path.
\begin{figure}
\centerline{\includegraphics[clip,width=10cm]{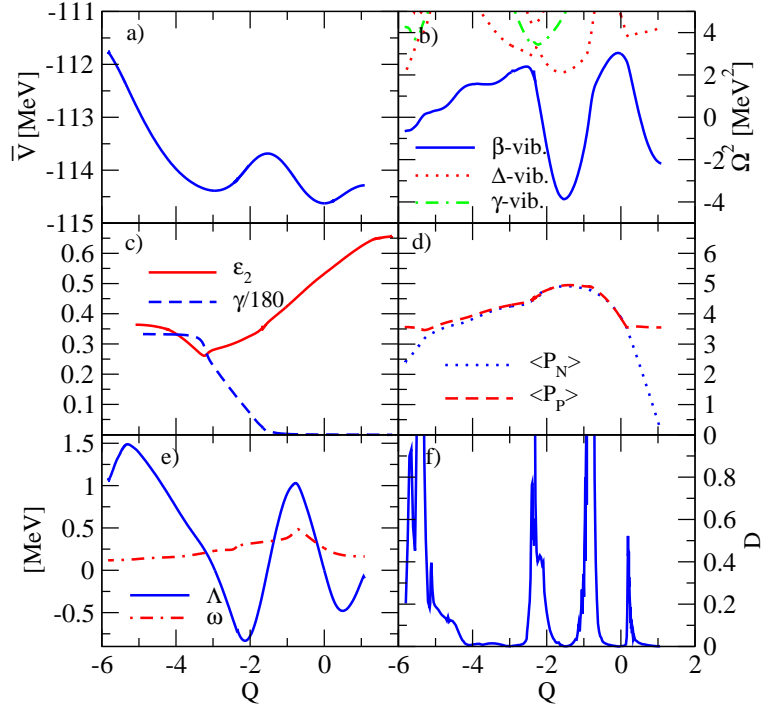}}
\caption{Large amplitude collective motion in $^{72}$Kr at I=2 following the 
  lowest RPA solution at equilibrium. a) Energy along the collective path.  b)
  The square of the lowest RPA frequencies. c) Deformation $\epsilon_2$ and 
  $\gamma /180^\circ$.  d) The dimensionless pairing
  operators $\left< P_\tau \right>$.  e) The cranking parameter
  $\Lambda$ and the rotational frequency $\omega$.  f) The decoupling
  measure, $D$.}
  \label{fig:kr72I2}
\end{figure}
\begin{figure}
\centerline{\includegraphics[clip,width=6cm]{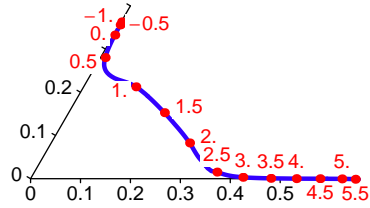}}
\caption{The deformation parameters in the $\epsilon_2-\gamma$ plane for 
  $^{72}$Kr at I=2. Labels denote the value of the collective coordinate.}
  \label{fig:kr72I2bg}
\end{figure}
In Fig.~\ref{fig:kr72I2} we see that if we start in the almost oblate
minimum and follow the collective path toward large oblate deformation
the situation is very similar as in the non-rotating case above. We
first see a decrease in the quadrupole moment but after an avoided
crossing with a pairing vibration the collective path changes its
nature and the collective path ends in a neutron pair-field
collapse. When we follow the collective path towards smaller oblate
deformation the system goes through an avoided crossing with a
$\gamma$-vibrational mode.The collective path will the go into the
triaxial plane and follow a path with approximately constant
$\epsilon_2$ deformation but with decreasing $\gamma$ from just below
$60$ to just above $0$. At this point the collective path goes through
another avoided crossing through which the computer code cannot find a
stable solution. By starting the calculation in the prolate minimum we
can follow the collective path to the same avoided crossing as as we
run into when starting in the oblate minimum. We can the add these two
path together to get the collective path from the oblate to the
prolate minimum and beyond.
\begin{figure}
  \centerline{\includegraphics[clip,width=8.0cm]{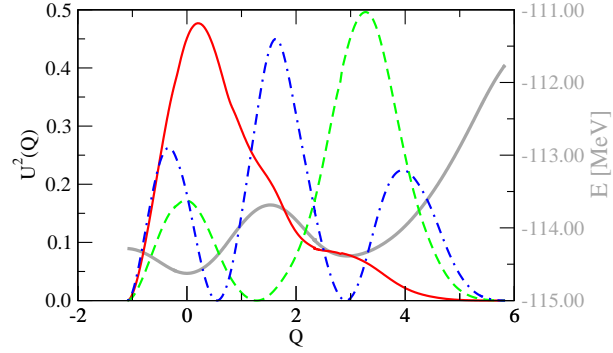}}
  \caption{The  wave function for the large amplitude collective
  motion in $^{72}$Kr at I=2, following the lowest RPA solution. 
The 
	first three $2^+$ eigen  functions are shown (as red solid, green
dashed and blue dashed-dotted line, respectively).
	These wave functions are periodic around $Q=0$ and $\pm 1.57$.
	The grey curve (with the scale on the 
	right-hand side) shows the collective potential energy for reference. }
 \label{fig:kr72I23D}
\end{figure}
When we solve the collective Schr\"odinger equation we get a ground
state in the oblate minimum which has a substantial tail into the
prolate minimum as can be seen in Fig.~\ref{fig:kr72I23D}. The first
excited state has its main peak in the prolate minimum with a second
peak in the oblate minimum.
\begin{table}
\caption{The deformation for the collective states as well as for the mean field 
	at $I=0$ and $2$ for $^{72}$Kr.}
\label{tab:kr72def} 
\begin{tabular}{|r|cccccccc|}
\hline
 $I_n$ & $0_1$ & $0_2$ & $0_3$ & $2_1$ & $2_2$ & $2_3$ & $0_{\rm MF}$ & $2_{\rm MF}$\\ 
\hline
 $\epsilon_2$ & 0.282& 0.368& 0.150& 0.302& 0.403& 0.355 & 0.321 & 0.334\\ 
 $\gamma$ & 60.0 & 0.0 & 0.0 & 42.4 & 9.8 & 19.9 & 60.0 & 59.4 \\ 
\hline
\end{tabular}
\end{table} 

\begin{figure}
  \centerline{\includegraphics[clip,width=8cm]{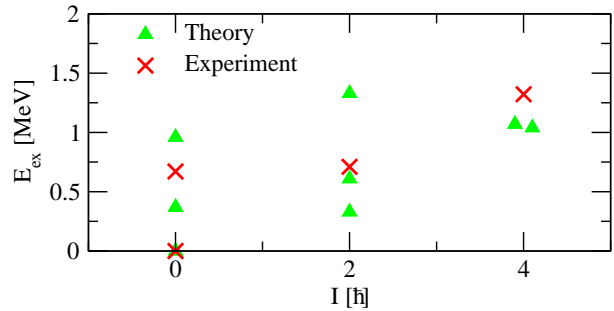}}
  \caption{Calculated and experimental~\cite{BM03} excitation spectrum for 
    $^{72}$Kr as a function of angular momentum.}
  \label{fig:kr72EI}
\end{figure}
Unfortunately, for the spectrum
of this nucleus, Fig.~\ref{fig:kr72EI}, our results suggest
that none of the states can be calculated reliably, due to the
relatively large values of $D$ close to the minimum. Indeed comparison
with the experimental data is extremely poor; a single collective
coordinate is simply insufficient in this case.

\subsection{$^{74}$Kr}
\begin{figure}
\centerline{\includegraphics[clip,width=10cm]{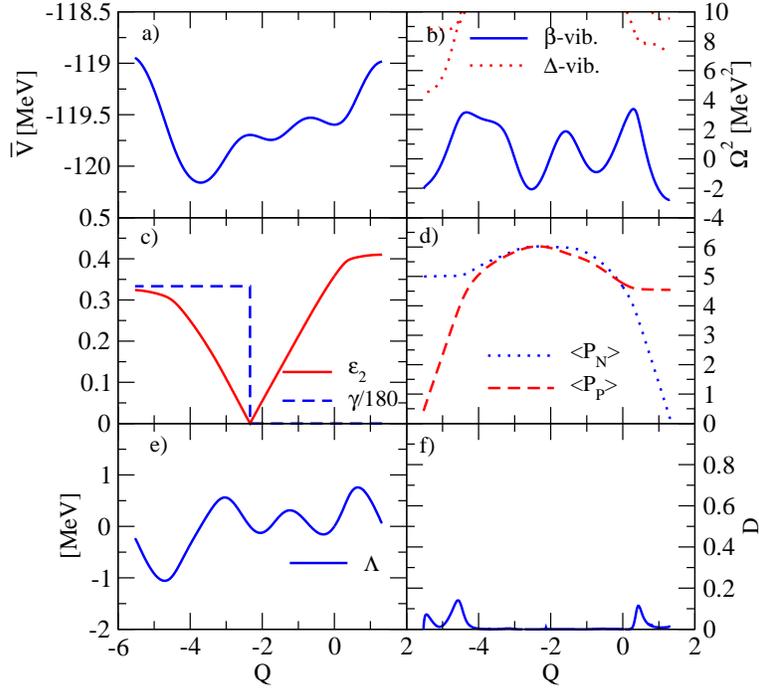}}
  \caption{Large amplitude collective motion in $^{74}$Kr at $I=0$
  following the lowest RPA solution at equilibrium. a) Energy along
  the collective path.  b) The square of the lowest RPA
  frequencies. c) Deformation $\epsilon_2$ and $\gamma /180^\circ$. d) The
  dimensionless pairing operators $\left< P_\tau \right>$.  e) The
  cranking parameter $\Lambda$.  f) The decoupling measure, $D$.
}
  \label{fig:kr74I0}
\end{figure}

In $^{74}$Kr the situation is similar to that of $^{72}$Kr. For zero
angular momentum the minimum has an oblate shape, as can be seen in
Fig.~\ref{fig:kr74I0}.
At large oblate deformations there is a collapse in the proton
pair-field proceeded by an avoided crossing between a
$\beta$-vibration and a pairing vibration that changes the direction
of the collective path. Towards smaller oblate deformation the
collective path goes over a spherical maximum and through two shallow
prolate minima before it ends in a state with collapsed neutron
pairing.
\begin{figure}
  \centerline{\includegraphics[clip,width=8.0cm]{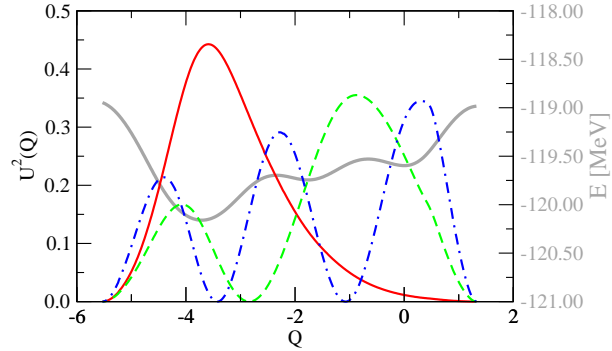}}
  \caption{The wave function for the large amplitude collective motion
  in $^{74}$Kr at $I=0$, following the lowest RPA solution. The first
  three $0^+$ eigen functions are shown (as red solid, green dashed
  and blue dashed-dotted line, respectively).  These wave functions
  are periodic around $Q=0$ and $\pm 1.57$.  The grey curve (with the
  scale on the right-hand side) shows the collective potential energy
  for reference. }  \label{fig:kr74I03D}
\end{figure}
The collective wave function, $U$, of the ground state is shown in
Fig.~\ref{fig:kr74I03D}.  It shows the wave function is dominated by
oblate shapes but contain a substantial tail of spherical and prolate
shapes. The first excited $0^+$ state is a mixture of a large prolate
component spread out over both the prolate minimum and a smaller
oblate part. The barrier between the two prolate minima is not enough
to build a collective state located mainly on each of these. The
second excited $0^+$ state is even more complicated and contain
prolate spherical and oblate shapes.

\begin{figure}
\centerline{\includegraphics[clip,width=10cm]{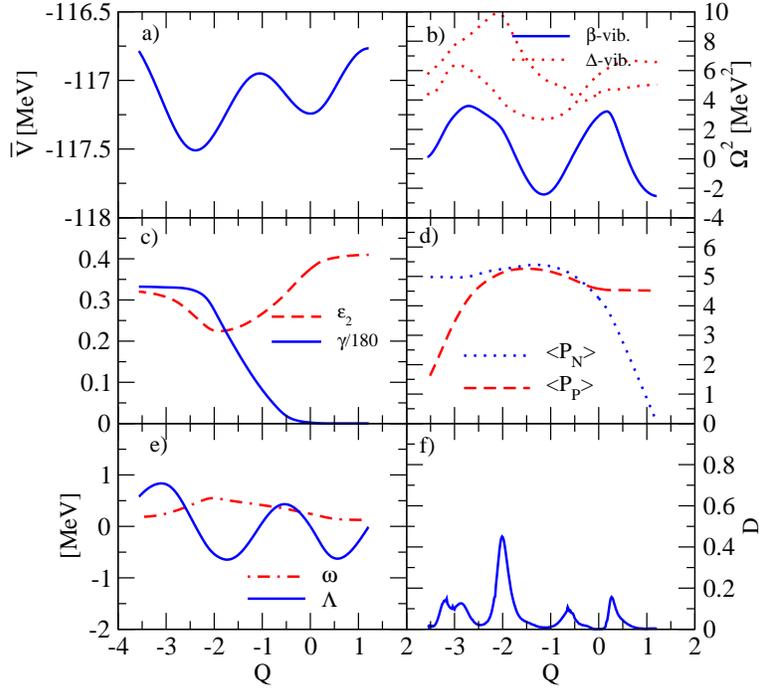}}
  \caption{Large amplitude collective motion in $^{74}$Kr at $I=2$ 
  following the lowest RPA
  solution at equilibrium). a) Energy along the collective path.  b)
  The square of the lowest RPA frequencies. c) Deformation $\epsilon_2$ and 
  $\gamma /180^\circ$.  d) The dimensionless pairing
  operators $\left< P_\tau \right>$.  e) The cranking parameter
  $\Lambda$ and the rotational frequency $\omega$.  f) The decoupling
  measure, $D$.}
  \label{fig:kr74I2}
\end{figure}

At $I=2$ we see in Fig.~\ref{fig:kr74I2} that the prolate and oblate
minimum are again connected via a path that is dominated by triaxial
shapes.
\begin{figure}
  \centerline{\includegraphics[clip,width=6cm]{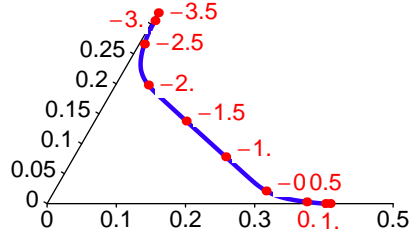}}
  \caption{The deformation parameters in the $\epsilon_2-\gamma$ plane for 
    $^{74}$Kr at $I=2$. Labels denote the value of the collective coordinate.}
  \label{fig:kr74I2bg}
\end{figure}
The collective path is ended by collapses in the pair-fields at large
oblate and prolate deformation. The prolate minimum we see at $I=2$ has
similar deformation as the second prolate minimum at $I=0$.
\begin{figure}
  \centerline{\includegraphics[clip,width=8.0cm]{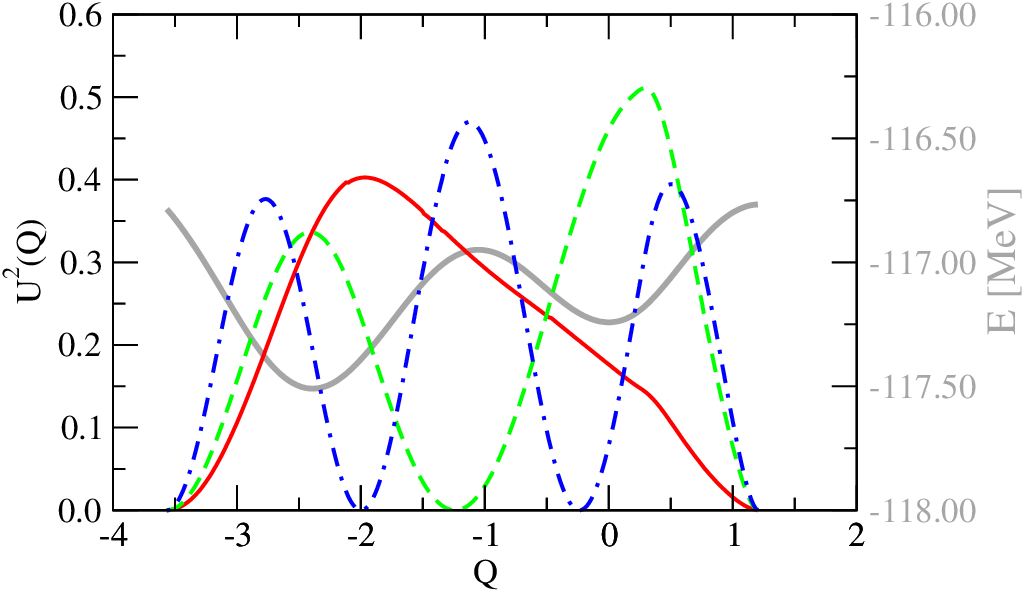}}
  \caption{The wave function for the large amplitude collective motion
  in $^{74}$Kr at $I=2$, following the lowest RPA solution. The first
  three $2^+$ eigen functions are shown (as red solid, green dashed
  and blue dashed-dotted line, respectively).  These wave functions
  are periodic around $Q=0$ and $\pm 1.57$.  The grey curve (with the
  scale on the right-hand side) shows the collective potential energy
  for reference.  } \label{fig:kr74I23D}
\end{figure}
The wave function of lowest $2^+$ state show a strongly collective
nature and even though the peak is situated in the oblate minimum it
has a strong admixture of triaxial and prolate shapes as can be seen
in Fig.~\ref{fig:kr74I23D}.  The wave function of higher-lying $2^+$
states contain a mixture of prolate, oblate and triaxial shapes.
\begin{table}
\caption{The deformation for the collective states as well as for the mean field 
	at $I=0$ and $2$ for $^{74}$Kr.}
\label{tab:kr74def} 
\begin{tabular}{|r|cccccccc|}
\hline
 $I_n$ & $0_1$ & $0_2$ & $0_3$ & $2_1$ & $2_2$ & $2_3$ & $0_{\rm MF}$ & $2_{\rm MF}$\\ 
\hline
 $\epsilon_2$ & 0.125& 0.135& 0.098& 0.304& 0.320& 0.314 & 0.213 & 0.259\\ 
 $\gamma$ & 60.0 & 0.0 & 0.0 & 49.1 & 43.7 & 46.3 & 60.0 & 57.9 \\ 
\hline
\end{tabular}
\end{table}

\begin{figure}
  \centerline{\includegraphics[clip,width=8cm]{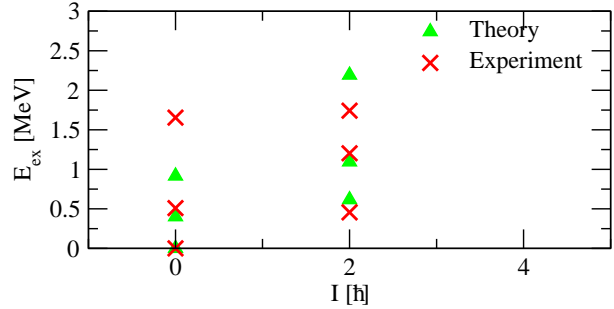}}
  \caption{Calculated and experimental~\cite{KC05} excitation spectrum for 
    $^{74}$Kr as a function of angular momentum.}
  \label{fig:kr74EI}
\end{figure}
In this case,  Fig.~\ref{fig:kr74EI}, we expect very good results for $0^+$
states since $D$ is small everywhere, and good, but slightly less accurate,
results for the $2^+$ states, since $D$ is slightly larger. 
Indeed, the results agree remarkably well with experimental data.

\subsection{$^{76}$Kr}
\begin{figure}
\centerline{\includegraphics[clip,width=10cm]{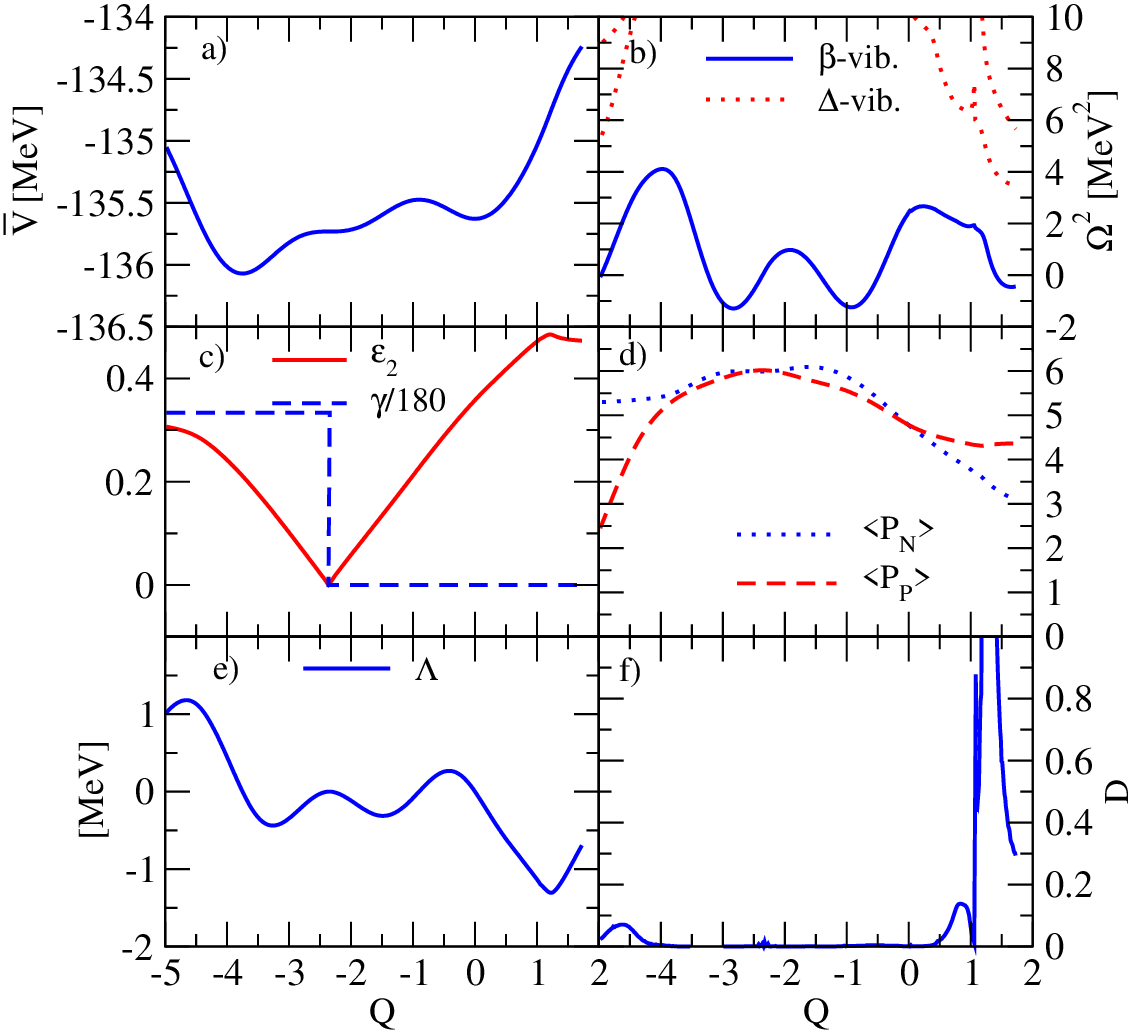}}
  \caption{Large amplitude collective motion in $^{76}$Kr at $I=0$
  following the lowest RPA solution at equilibrium. a) Energy along
  the collective path.  b) The square of the lowest RPA
  frequencies. c) Deformation $\epsilon_2$ and $\gamma /180^\circ$. d) The
  dimensionless pairing operators $\left< P_\tau \right>$.  e) The
  cranking parameter $\Lambda$.  f) The decoupling measure, $D$.
}
  \label{fig:kr76I0}
\end{figure}

In $^{76}$Kr the collective path is almost identical to that of $^{74}$Kr. For the 
unrotated case the lowest minimum has an oblate shape as can bee seen in
Fig.~\ref{fig:kr76I0}. 
Towards small oblate deformation the collective path goes over a
spherical maximum and through two shallow prolate minima. At each end
of the collective path there is a strong reductions in the
pair-fields. Due to numerical difficulties of finding a converged
solution we are not able continue the calculation all the way to the
pairing collapsed state.
\begin{figure}
  \centerline{\includegraphics[clip,width=8.0cm]{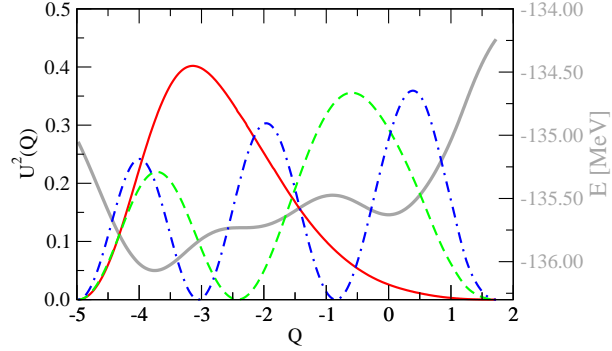}}
  \caption{The  wave function for the large amplitude collective
  motion in $^{76}$Kr at $I=0$, following the lowest RPA solution. The first
  three $0^+$ eigen functions are shown (as red solid, green dashed
  and blue dashed-dotted line, respectively).  These wave functions
  are periodic around $Q=0$ and $\pm 1.57$.  The grey curve (with the
  scale on the right-hand side) shows the collective potential energy
  for reference. } \label{fig:kr76I03D}
\end{figure}
The collective wave function, $U$, of the ground state is shown in
Fig.~\ref{fig:kr76I03D}.  It shows the wave function is dominated by
oblate shapes but contains a substantial tail of spherical and prolate
shapes. The first excited $0^+$ state is a mixture of a large prolate
component and a smaller oblate part. The second excited $0^+$ state is
even more complicated and mixes prolate spherical and oblate shapes.

\begin{figure}
\centerline{\includegraphics[clip,width=10cm]{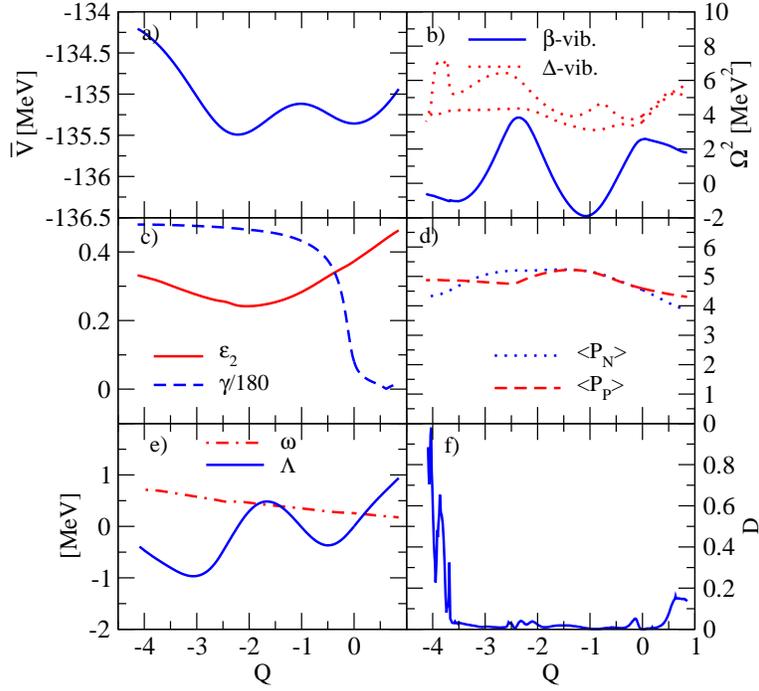}}
  \caption{Large amplitude collective motion in $^{76}$Kr at $I=2$
  following the lowest RPA solution at equilibrium. a) Energy along
  the collective path.  b) The square of the lowest RPA
  frequencies. c) Deformation $\epsilon_2$ and $\gamma /180^\circ$. d) The
  dimensionless pairing operators $\left< P_\tau \right>$.  e) The
  cranking parameter $\Lambda$ and the rotational frequency $\omega$.
  f) The decoupling measure, $D$.}  
  \label{fig:kr76I2}
\end{figure}

At $I=2$ we see in Fig.~\ref{fig:kr76I2} that the prolate and oblate
minimum are connected via a path that is dominated by triaxial
shapes. The lowest minimum is not oblate at all but instead strongly
triaxial ($\gamma \approx 90^\circ$). This is due to the fact that the
collective path goes from the triaxial minimum rapidly through an
oblate shaped state to a secondary prolate minimum.
\begin{figure}
  \centerline{\includegraphics[clip,width=6cm]{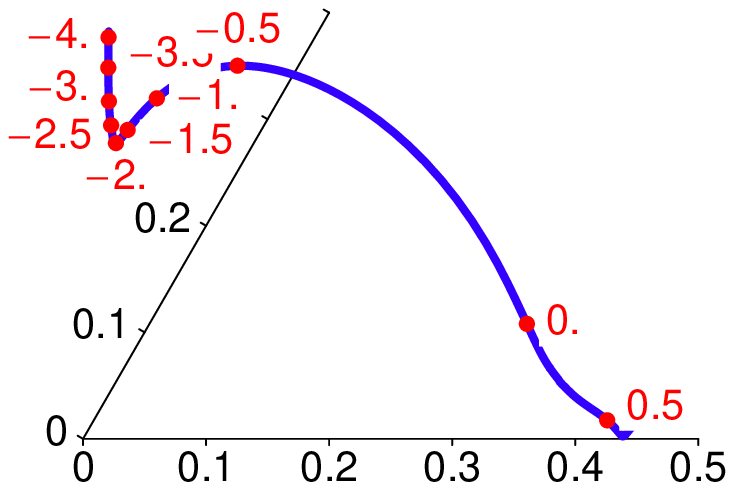}}
  \caption{ The deformation parameters in the $\epsilon_2-\gamma$ plane for
    $^{76}$Kr at $I=2$. Labels denote the value of the collective coordinate.}
  \label{fig:kr76I2bg}
\end{figure}
The collective path ends due to numerical difficulties at relatively small prolate  
and triaxial deformations. 
\begin{figure}
  \centerline{\includegraphics[clip,width=8.0cm]{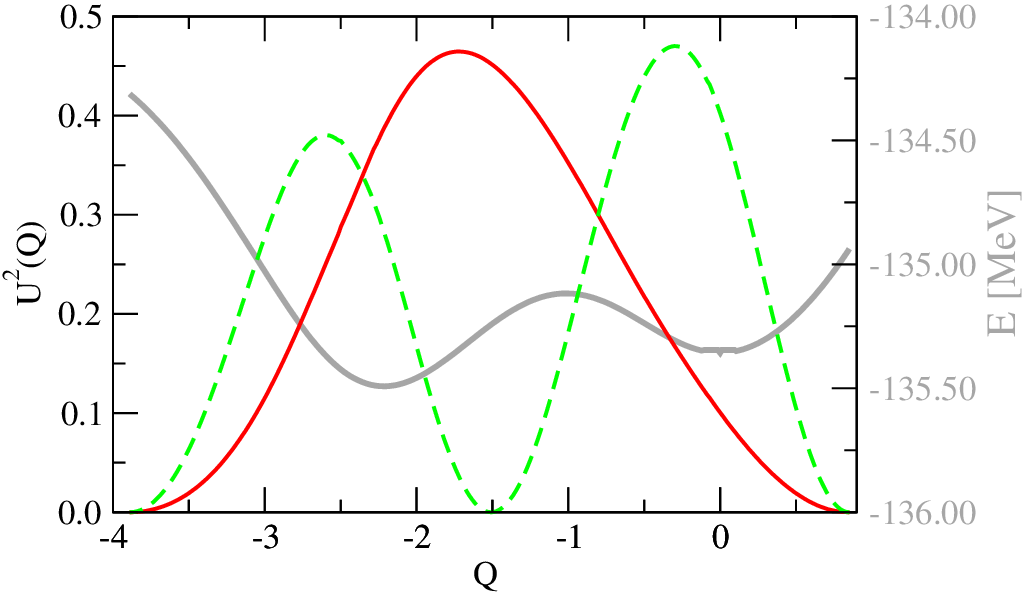}}
  \caption{The  wave function for the large amplitude collective
  motion in $^{76}$Kr at $I=2$, following the lowest RPA solution. The first
  three $2^+$ eigen functions are shown (as red solid, green dashed
  and blue dashed-dotted line, respectively).  These wave functions
  are periodic around $Q=0$ and $\pm 1.57$.  The grey curve (with the
  scale on the right-hand side) shows the collective potential energy
  for reference. } \label{fig:kr76I23D}
\end{figure}
The wave function of the lowest $2^+$ state shows a strongly collective
nature and even though the peak is situated in the triaxial minimum it
has a strong admixture of oblate and prolate shapes as can be seen in
Fig.~\ref{fig:kr74I23D}.  The wave functions of higher lying $2^+$
states contain a mixture of prolate, oblate and triaxial shapes. The
energy of the states can only be seen as an upper limit since we have
not been able to follow the collective path very far in this case.
\begin{table}
\caption{The deformation for the collective states as well as for the mean field 
	at $I=0$ and $2$ for $^{76}$Kr.}
\label{tab:kr76def} 
\begin{tabular}{|r|cccccccc|}
\hline
 $I_n$ & $0_1$ & $0_2$ & $0_3$ & $2_1$ & $2_2$ & $2_3$ & $0_{\rm MF}$ & $2_{\rm MF}$\\
\hline
 $\epsilon_2$ & 0.048& 0.150& 0.142& 0.253& 0.266& 0.257& 0.211 & 0.244\\ 
 $\gamma$ & 60.0 & 0.0 & 0.0 & 35.3 & 28.1 & 34.0 & 60.0 & 56.0 \\ 
\hline
\end{tabular}
\end{table}

\begin{figure}
  \centerline{\includegraphics[clip,width=8cm]{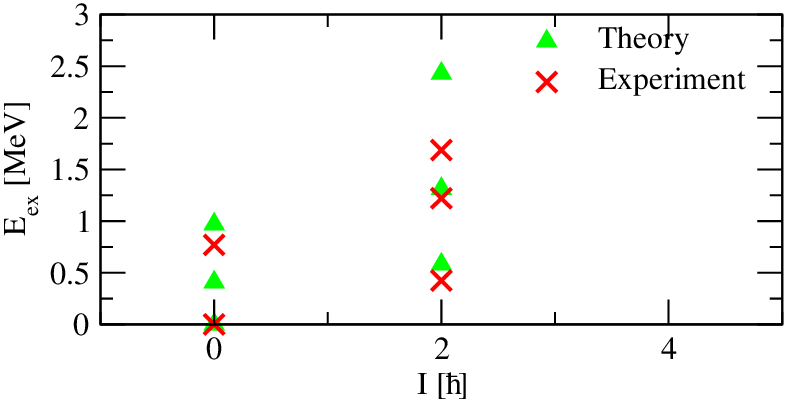}}
  \caption{Calculated and experimental~\cite{KC05} excitation spectrum for 
    $^{76}$Kr as a function of angular momentum.}
  \label{fig:kr76EI}
\end{figure}
Thus, see Fig.~\ref{fig:kr76EI}, the energies first two $0^+$
states should be reliable, but the third is somewhat more doubtful,
due to the peak in $D$ at $Q=-1$. The $2^+$ states are of similar 
quality. It is thus surprising that theory predicts a low-lying second
$0+$-state that has not been observed in experiment!

\subsection{$^{78}$Kr}
\begin{figure}
  \centerline{\includegraphics[clip,width=8.0cm]{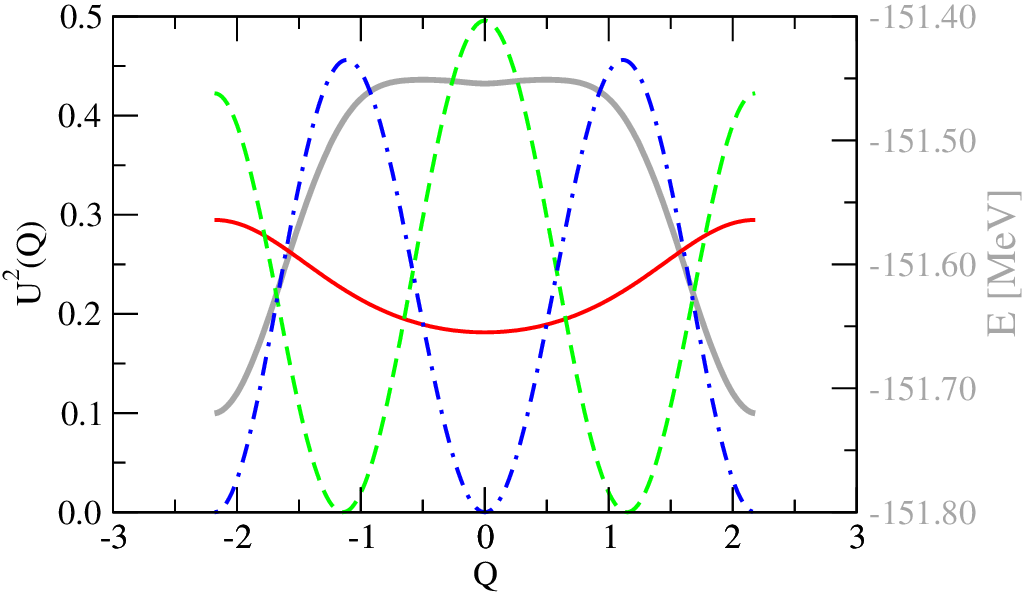}}
  \caption{The  wave function for the large amplitude collective
  motion in $^{78}$Kr at $I=0$, following the lowest RPA solution. The first
  three $0^+$ eigen functions are shown (as red solid, green dashed
  and blue dashed-dotted line, respectively).  These wave functions
  are periodic around $Q=0$ and $\pm 1.57$.  The grey curve (with the
  scale on the right-hand side) shows the collective potential energy
  for reference. } \label{fig:kr78I01D}
\end{figure}

For $^{78}$Kr we have behaviour similar to $^{68}$Se in that the
lowest RPA mode in the minimum is a $\gamma$-vibration. The prolate
and oblate minimum are therefore connected through triaxial states.
\begin{figure}
\centerline{\includegraphics[clip,width=10cm]{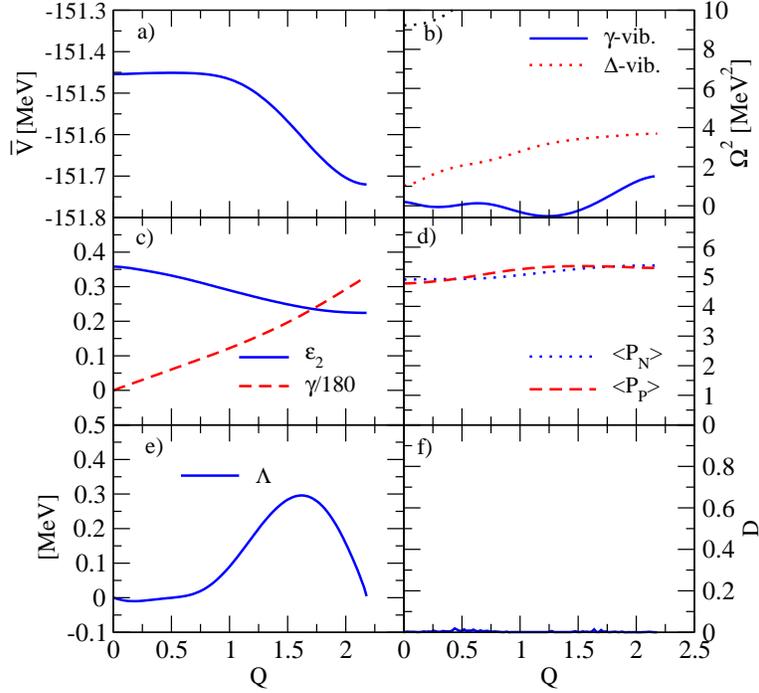}}
  \caption{Large amplitude collective motion in $^{78}$Kr at $I=0$
  following the lowest RPA solution at equilibrium. a) Energy along
  the collective path.  b) The square of the lowest RPA
  frequencies. c) Deformation $\epsilon_2$ and $\gamma /180^\circ$.  d) The
  dimensionless pairing operators $\left< P_\tau \right>$.  e) The
  cranking parameter $\Lambda$.  f) The decoupling measure, $D$.}
  \label{fig:kr78I0}
\end{figure}
\begin{figure}
  \centerline{\includegraphics[clip,width=6cm]{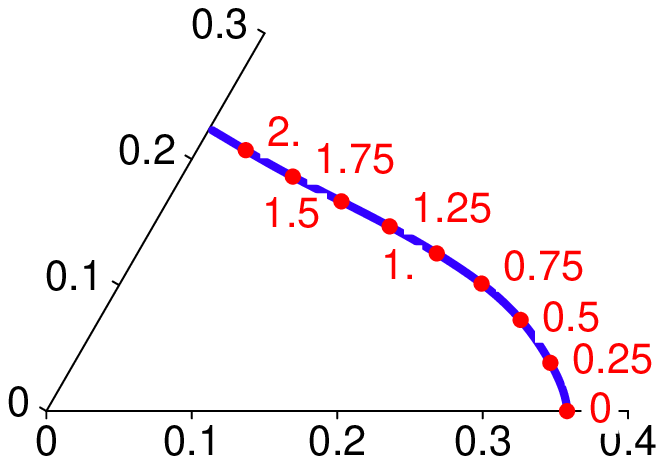}}
  \caption{The deformation parameters in the $\epsilon_2-\gamma$ plane for 
    $^{78}$Kr at $I=0$. Labels denote the value of the collective coordinate.}
  \label{fig:kr78I0bg}
\end{figure}
The potential barrier between the oblate minimum, which is lowest in
energy, and the secondary prolate minimum is only $250$ keV as can be
seen in Fig.~\ref{fig:kr78I0}. In Fig.~\ref{fig:kr78I01D} we see that
this leads to a situation very similar to that of $^{68}$Se where the
collective ground state wave function, $U$, represent a approximately
equal mixture of all possible triaxial shapes with an approximately
constant $\epsilon_2$-deformation and pair-fields.
We also see that the excitation spectrum in $^{78}$Kr is very similar
to that of $^{68}$Se. We have two almost degenerate excited $0^+$
states that can be interpreted as rotational excitation (in
$\gamma$). The fact these excited states lie at $1$ MeV in
$^{78}$Kr compared to $2$ MeV $^{68}$Se in simply due to the longer
'distance' between the prolate and the oblate minimum in $^{78}$Kr.

\begin{figure}
\centerline{\includegraphics[clip,width=10cm]{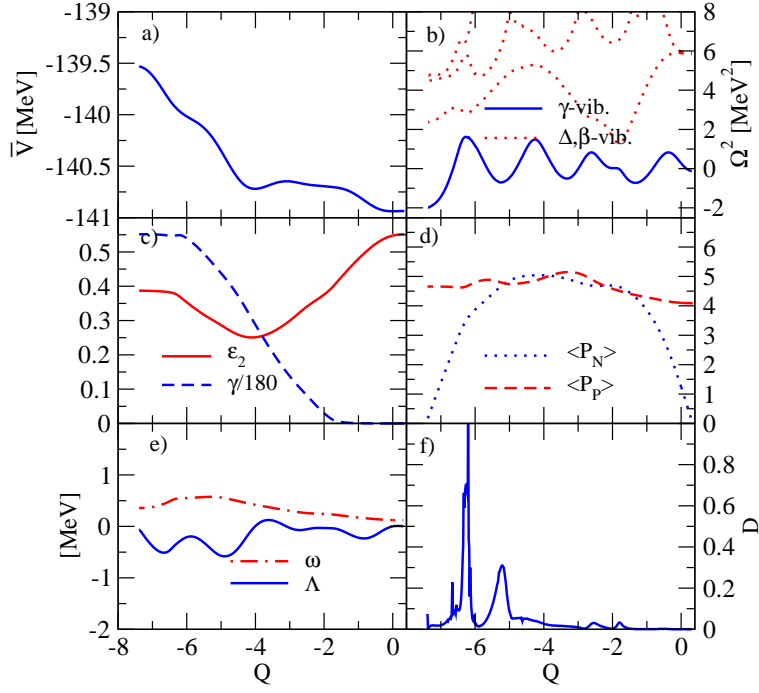}}
  \caption{Large amplitude collective motion in $^{78}$Kr at $I=2$
  following the lowest RPA solution at equilibrium). a) Energy along
  the collective path.  b) The square of the lowest RPA
  frequencies. c) Deformation $\epsilon_2$ and $\gamma /180^\circ$. d) The
  dimensionless pairing operators $\left< P_\tau \right>$.  e) The
  cranking parameter $\Lambda$ and the rotational frequency $\omega$.
  f) The decoupling measure, $D$.}  
\label{fig:kr78I2}
\end{figure}

\begin{figure}
  \centerline{\includegraphics[clip,width=8.0cm]{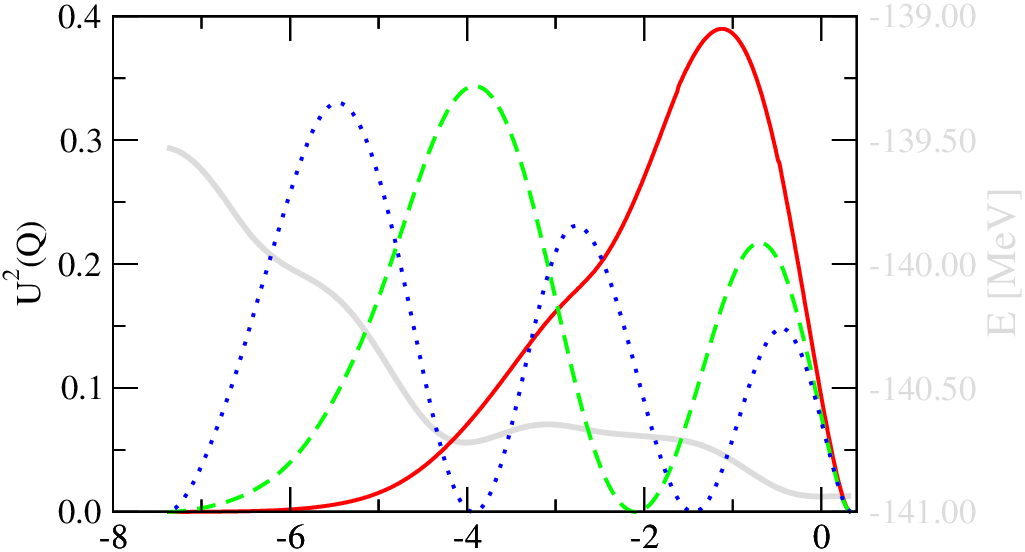}}
  \caption{The wave function for the large amplitude collective
  motion in $^{78}$Kr at $I=2$, following the lowest RPA solution. 
The first
  three $2^+$ eigen functions are shown (as red solid, green dashed
  and blue dashed-dotted line, respectively).  These wave functions
  are periodic around $Q=0$ and $\pm 1.57$.  The grey curve (with the
  scale on the right-hand side) shows the collective potential energy
  for reference. 
} \label{fig:kr78I23D}
\end{figure}

At at finite rotational frequency ($I=2$) we have a slightly different
picture for $^{78}$Kr. The lowest RPA solution in the almost axial
prolate minimum is a neutron paring vibration. This leads to a rapid
collapse of the neutron pair-field at small negative values of the
collective coordinate, as can be seen in Fig.~\ref{fig:kr78I2}.
\begin{figure}
\centerline{\includegraphics[clip,width=6cm]{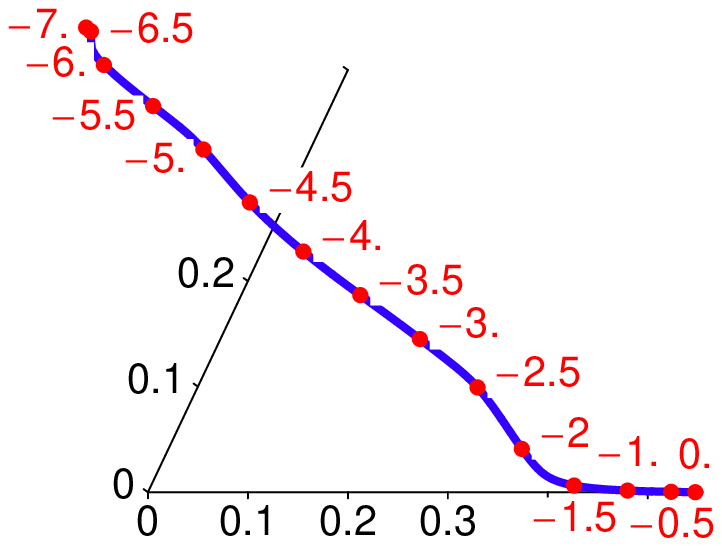}}
\caption{The deformation parameters in the $\epsilon_2-\gamma$ plane for 
  $^{78}$Kr at $I=2$. Labels denote the value of the collective coordinate.}  
\label{fig:kr78I2bg}
\end{figure}
At positive values of $Q$ there is an increase in the neutron
pair-field and a decrease in the $\epsilon_2$-deformation. At
$Q\approx 1.8$ there is an avoided crossing that changes the direction
of the collective path into the triaxial plane.  The path goes towards
larger $\gamma$ until we reach a almost oblate state with $\gamma
\approx 60$ and $Q \approx 4.1$. The collective path passes through
this secondary minimum in to triaxial plane. At $Q\approx 6.2$ we have
a second avoided crossing and a transition into a path with decreasing
neutron pair-field which leads to a neutron paring collapse at $
Q\approx 7.4$.  Due to the shallowness of the barrier between the
prolate and oblate sates the ground state wave function has a
substantial tail stretching out of the prolate minimum through the
triaxial plane into the oblate secondary minimum, see
Fig.~\ref{fig:kr78I23D}.
There are another two low lying $2^+$ states. The first of these is
dominated by a oblate component but has also a sizable prolate part.
\begin{table}
\caption{The deformation for the collective states as well as for the
	mean field at $I=0$ and $2$ for $^{78}$Kr.}
\label{tab:kr78def} 
\begin{tabular}{|r|cccccccc|}
\hline
 $I_n$ & $0_1$ & $0_2$ & $0_3$ & $2_1$ & $2_2$ & $2_3$ & $0_{\rm MF}$ & $2_{\rm MF}$\\
\hline
 $\epsilon_2$ & 0.266& 0.276& 0.278& 0.397& 0.286& 0.266 & 0.224 & 0.548\\ 
 $\gamma$ & 26.3 & 20.8 & 24.2 & 7.1 & 32.3 & 52.5 & 60.0 & 60.0 \\ 
\hline
\end{tabular}
\end{table}

\begin{figure}
  \centerline{\includegraphics[clip,width=8cm]{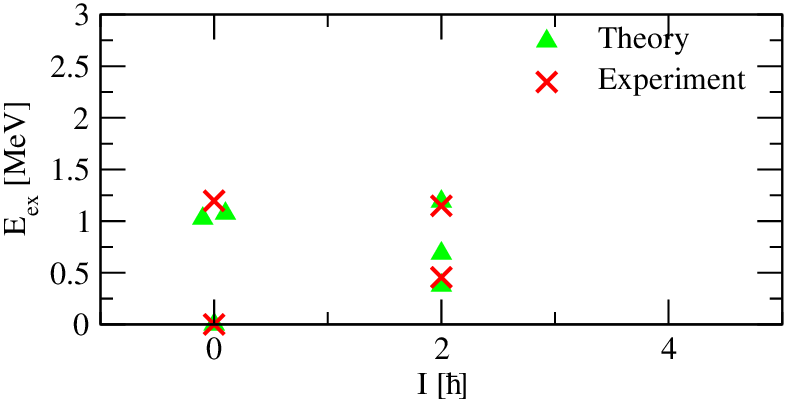}}
  \caption{Calculated and experimental~\cite{KC05} excitation spectrum for 
    $^{78}$Kr as a function of angular momentum.}
  \label{fig:kr78EI}
\end{figure}
In this case, Fig.~\ref{fig:kr78EI}, the spectrum is somewhat
similar to $^{68}$Se, due to the extreme $\gamma$-soft character of the
$0^+$ states. In this case one of the two excited $0^+$ states has
actually been detected. The data obtained in these difficult
experiments are not of sufficient quality to rule out 
that this is actually a narrow doublet.
The data on $2^+$ states is also compatible with an
additional low-lying $2^+$ that has not been seen in experiment.

\section{Conclusions}
\label{sec:Conclusions}
We have extended the method of calculating the self-consistent
collective path presented in~\cite{DK00,AW04} to include constraints
on angular momentum and particle number and implemented it for the
quadrupole+pairing Hamiltonian~\cite{BK68}. The method consists of
finding a series of points fulfilling the force equation, where the
local direction of the collective path is determined in each point by
the local normal modes. The local RPA equations and the force equation
are solved in a double iterative process with constraints on angular
momentum and particle numbers along the collective path. The method
allows us to determine the collective coordinate from the Hamiltonian
without having to assume a priori which are the relevant degrees of
freedom.

The results confirm the importance of pairing collapsed states for the
collective path as suggested in~\cite{AW04}. We have also seen a very
different behaviour of the collective path between the prolate and
oblate minimum depending on $N$ and $Z$ as well as weather we are
looking at the ground state or an rotating state. Without rotation the
path either goes through a spherical maximum or through the triaxial
plane. In the rotating case the two minima are always connected via
the triaxial plane. This difference comes because at finite angular
momentum the system tries to avoid the spherical state because the
cost of generating angular momentum gets larger as the deformation
gets smaller. This effect would not have been seen so clearly if the
calculation was done at constant rotational frequency instead of at
constant angular momentum. The difference would be less pronounced if
we would have allowed for more the one collective coordinate.

The results shed new light on the behaviour of these nuclei, and make
predictions which can be of importance experimentally, especially the
extreme $\gamma$-soft nature of $^{68}$Se and $^{72}$Kr at low angular
momentum. Discovery of the predicted doublets of excited $0^+$ states
would shed light on the nature of shape mixing in these nuclei.  It is
extremely encouraging that where we expect our method to be reliable
(i.e., where $D$ is small) we get a good correspondence with
experiment, and where we have a poor correspondence, decoupling is
poor. We have also seen that the nature of shape-mixing and
shape-coexistence is indeed quite complicated in the group of nuclei
discussed here, and changes rapidly along the isotope chain. The
picture seems to be less systematic than discussed in the
literature. Mixing is large in all cases.

In this paper we have implemented a method to find the adiabatic
self-consistent collective path for a nuclei. A technique to truncate
the basis in which the RPA equations are solved has been improved and
a good agreement between the full and truncated RPA is found. To solve
the RPA equations in a limited basis has proven to be a useful and
practical way of calculating the collective path within the local
harmonic approximation. It remains to be investigated how we can
include the covariant terms in the RPA in a suitable approximation.

\section*{Acknowledgement}
This work was supported by the UK Engineering and Physical Sciences Research 
Council (EPSRC)

\appendix
\section{Comparison with Matsuo {\em et al.}} 
In the work of Matsuo {\em at al.}, the adiabatic large amplitude
collective motion (ALACM) is described in what looks superficially
entirely different than our method.  Nonetheless, the final results
are so similar that there must be similarities. It is therefore useful
to analyse the problem in various ways and compare the two methods in
a unified notation. The key difference is the direct formulation in
terms of mean field equations by Matsuo \emph{et al.}, and our
insistance on classical mechanics.

Let us initially look at one way to derive the localised harmonic
approach in our notation, which we shall than tie closely to the work
in Ref.~\cite{MN00}. We expand the collective energy (actually the
generalised density matrix) to second order in small fluctuations,
thus obtaining a Hamiltonian. 
The method is based on ``removing the linear
terms'', which means that at any point $P$, which is really a
time-even HFB state, along the collective path, we approximate the
Hamiltonian by the quadratic form
\begin{equation}
{\cal H}= \frac{1}{2} \pi_\alpha B^{\alpha\beta}(P) \pi_\beta+V(P)+V_{,\alpha}(P) \xi^\alpha +\frac{1}{2} V_{,\alpha\beta}(P) \xi^\alpha
\xi^\beta\quad.
\end{equation}
In order to look at the local fluctuations, we need to remove the
term linear in $\xi$. We first modify $\cal H$ by constraints (such as those
for particle number and angular momentum; we shall include particle
number only here). With the Lagrange multiplier term $\mu(N-N_0)$ we really 
have
\begin{equation}
{\cal H} -\lambda N = \frac{1}{2} \pi_\alpha (B^{\alpha\beta}-\mu \partial_{\pi_\alpha} \partial_{\pi_\beta}N)(q_0) \pi_\beta+V(q_0)+
(V_{,\alpha}-\mu N_{,\alpha})(q_0) \xi^\alpha +\frac{1}{2} (V_{,\alpha\beta}(q_0)-\mu N_{,\alpha\beta}-\mu_{,\alpha}
N_{,\beta}-\mu_{,\beta}N_{,\alpha}) \xi^\alpha \xi^\beta\quad.
\end{equation}
In order to remove the linear term, we require that it can be removed
by purely collective term, i.e., a term of the form $\lambda Q$, with
$\lambda$ defined at the collective point $Q=q_0$.  We thus require
that
\begin{equation}
V_{,\alpha}-\mu N_{,\alpha}=\lambda f_{,\alpha}\quad,
\end{equation}
with $f_{,\alpha}=Q_{,\alpha}$, and we get the quadratic
Hamiltonian (where we have absorbed the terms arising from expanding
$\langle H-\mu N \rangle$ for fixed $\mu$ in $B$ and $V$, and denoted
the compound object by a tilde). The Hamiltonian with the constraint 
$\lambda Q$ becomes
\begin{equation}
{\cal H}'=\frac{1}{2} \pi_\alpha ({\tilde B}^{\alpha\beta}) \pi_\beta+V
+\frac{1}{2} ({\tilde V}_{,\alpha\beta}-\mu_{,\alpha}
N_{,\beta}-\mu_{,\beta}N_{,\alpha}-\lambda f_{,\alpha\beta})
\xi^\alpha \xi^\beta\quad.
\end{equation}
It has been argued coherently in Ref.~\cite{DK00} that the final term
is equivalent to the covariant term in the derivative
\[ \Gamma_{\alpha\beta}^\gamma V_{,\gamma},\] 
and we are now left with a simple normal mode problem, which is of
course nothing more than the usual RPA. The self-consistency is now
obtained by requiring $f$ to be one of the normal mode
directions.
We thus can write the RPA as two coupled equation (or as a single one)
\footnote{$f_{,\alpha}=f^{1}_{,\alpha}=\partial_{\xi^\alpha}Q$ and $g^{\alpha}=g^{\alpha}_{,1}=\partial_{\pi_\alpha}P$
in the notation of Ref.~\cite{DK00}.}
\begin{align}
{\tilde B}^{\alpha\beta} f_{,\beta}&=\bar{B} g^{\alpha}\quad,
\label{eq:apcRPA1}\\
({\tilde V}_{,\alpha\beta}-\mu_{,\alpha}
N_{,\beta}-\mu_{,\beta}N_{,\alpha}-\lambda f_{,\alpha\beta})g^\beta&=
\bar{V}_{;QQ} f_\alpha\quad.
\label{eq:apcRPA2}
\end{align}

Let us now compare this approach with that of \cite{KN03,KN05}, which
is cast in the form of a set of variational equations. If we take into
account that to first order the time-even part of fluctuations is
$\xi^\alpha/\sqrt{2}$, with $\pi_\alpha/\sqrt{2}$ the time-odd part,
it is relatively straight-forward to show that the three equations
(2.34-2.36) in ref.~\cite{MN00} would be written as\\
(\textbf{note:} 
\begin{itemize}
\item we use $\mu$ where Ref.~\cite{MN00} uses $\lambda$; 
\item in our notation
$\lambda=\bar{V}(Q)_{,Q}$, where we use the bar to denote a collective
quantity, as distinguished from the quantity defined in the full space
\cite{DK00}.
\item $f_{,\alpha}=\sqrt{2} \delta \langle \hat Q \rangle$.
\item $g^{,\alpha}=\sqrt{2}\delta \langle \hat P \rangle$.)
\end{itemize}
\begin{align}
V_{,\alpha} -\mu N_{,\alpha}-\lambda f_{,\alpha}&=0\quad,\\
{\tilde B} ^{\alpha\beta} f_{,\beta}-\bar{B}g^{\alpha}&=0\quad,\\
\frac{1}{2}\left[f_{,\alpha} {\tilde B}^{\alpha\beta}f_{,\beta}\right]_{,\gamma}-\bar{B}
\frac{1}{2}\left[f_{,\gamma\beta}g^\beta +\bar{\Gamma} f_{,\gamma}\right]&=0\quad.
\end{align}
The first of these relations is the one that removes the linear term, as above.
The second defines the collective momentum, and finally the last equation
defines the affine connection $\bar{\Gamma}$,
\begin{equation}
\bar B \bar \Gamma= \frac{1}{2}\tilde{B}^{\alpha\beta}_{,\gamma} f_{,\alpha}f_{,\beta} g^\gamma
\quad.
\end{equation}
Exactly the same relation can be found in Ref.~\cite{DK00}.
Whereas in the local fluctuation approach we construct fluctuations in
all directions, the work in Ref.~\cite{MN00} differentiates only with
respect to the collective coordinates. This implies an assumption
about block-diagonality of $\Gamma$ that is justified for exact decoupling.
We get for Eq.~(3.1) in Ref.~\cite{MN00},
\begin{equation}
(V_{,\alpha\beta}-\mu N_{,\alpha\beta})g^\beta-\bar{V}_{;QQ} f_{,\alpha}
-\bar{V}_{,Q} \left[f_{,\alpha\beta}g^\beta +\bar{\Gamma} f_{,\alpha}\right]
-\mu_{,Q} N_{,\alpha}=0
\end{equation}
This can also be rewritten as (using $\bar{V}_{,Q}=\lambda$ and $\mu_{,Q}=
\mu_{,\alpha} g^{\alpha}$)
\begin{equation}
(V_{,\alpha\beta}-\mu N_{,\alpha\beta}-\lambda f_{,\alpha\beta}-\mu_{,\beta}N_{,\alpha})g^\beta
=
\lambda_{,Q} f_{,\alpha}, \label{eq:apV3}
\end{equation}
which is almost what we would argue constitutes the covariant RPA
equation, cf.~Eqs.~(\ref{eq:apcRPA2}).
There is a very subtle difference, and that is due to the fact
that the matrix on the left-hand-side in \ref{eq:apV3} is not symmetric
under interchange of the labels $\alpha$ and $\beta$. The term that
is missing as compared to the local fluctuation approach discussed
above is $-\mu_{,\alpha}N_{,\beta}g^\beta=-\mu_{,\alpha} N_{,Q}$.
The absence comes from treating $\mu=\mu(Q)$ in the approach
of Ref.~\cite{MN00} as an external parameter, whereas, in the
language of that paper, we take $\mu(\langle \hat Q \rangle)$, 
i.e., as a function of the mean-field expectation value
of $\hat Q$, which clearly changes with the mean field, and thus
leads to the additional term.

We believe that symmetry of all the RPA equations is crucial, and thus
are worried about such missing terms.  Of course, in the current paper
we have grossly neglected all covariant terms, which clearly is also
objectionable.  We do wonder, however, if this lack of symmetry is
what leads to the $K$-symmetry non-preserving path seen in
Ref.~\cite{KN05}.  Without angular momentum constraints, the local RPA
for axial states will \emph{always} have eigenvalues with definite
$\Delta K$, i.e., to change from following a $\beta$ to a $\gamma$
vibration, we must have a real rather than avoided crossing in the
RPA.

The further developments in Ref.~\cite{MN00} are also interesting;
they correspond to trying to evaluate $f_{,\alpha\beta}$, the
variation of the collective coordinate operator. Even if it were a one
body operator, as we tacitly assume, and is necessary for exact
decoupling, the equations only fix the 2 quasi-particle part of this
operator, i.e., terms going like $c^\dagger c^\dagger$ or $cc$. The
approximation made in Ref.~\cite{MN00} is to assume that the
$c^\dagger c$ matrix elements are identically zero. This is the only
closed form approximation that can be made, but must be dealt with
carefully: it would fail for some problems with exact decoupling.  In
an approach that uses a basis of operators, one could evaluate such
terms by the relevant matrix elements of the operators, multiplied by
their coefficients \cite{DK00}.

We thus conclude that the underlying expressions in our approach and
that set out in Ref.~\cite{MN00} are almost, and should be completely,
identical. We are concerned about the missing term in the potential
expansion.

\end{document}